\documentclass[12pt]{article}
\usepackage[left=2cm, right=2cm, bottom=2cm ,top = 2cm, footskip=1cm]{geometry}
\usepackage[font=small,labelfont=bf]{caption}

\usepackage{amsmath}

\usepackage{setspace}
\usepackage{lineno}

\let\counterwithin\relax
\usepackage{chngcntr}

\usepackage{hyperref}
\usepackage{rotating}
\usepackage[utf8]{inputenc}
\usepackage[T1]{fontenc}

\usepackage{array, booktabs, makecell, hhline}
\setcellgapes{5pt}

\usepackage[authoryear,sort&compress]{natbib}
\setcitestyle{open={(},close={)}}

\usepackage{enumitem}

\usepackage{graphicx}
\usepackage{amssymb}

\usepackage{float} % figure placings

\usepackage[section]{placeins}

\title{Seizure duration is associated with multiple timescales in interictal iEEG band power}

\author{Mariella Panagiotopoulou$^{1*}$, Gabrielle M. Schroeder$^1$,\\ Jess Blickwedel$^1$, Fahmida A Chowdhury$^3$, Beate Diehl$^3$, Jane de Tisi$^{3}$\\ John S. Duncan$^3$, Alison Cronie$^4$, Jennifer Falconer$^4$, Ryan Faulder$^1$,\\ Veronica Leach$^4$, Shona Livingstone$^4$, \\Rhys H. Thomas$^2$, Peter N. Taylor$^{1,2,3}$, \\ Yujiang Wang$^{1,2,3**}$}

\begin{document}

\maketitle

% author affiliations
\begin{enumerate}
\item{CNNP Lab (www.cnnp-lab.com), Interdisciplinary Computing and Complex BioSystems Group, School of Computing, Newcastle University, Newcastle upon Tyne, United Kingdom}
\item{Faculty of Medical Sciences, Newcastle University, United Kingdom}
\item{UCL Queen Square Institute of Neurology, Queen Square, London, United Kingdom}
\item{NHS Greater Glasgow and Clyde, Glasgow, United Kingdom}
\end{enumerate}

\subsection*{Key Points} % Fewer than 140 characters
\begin{itemize}
    \item Seizure duration may be modulated by multiple cyclical patterns as revealed from interictal iEEG band power.
    \item A combination of cycles on different timescales could explain the diversity in seizure duration in individual subjects. 
\end{itemize}

\begin{itemize}[label={}]
\item $^*$ Mariella Panagiotopoulou (ORCID ID: 0000-0002-4047-9170)\\
Urban Sciences Building, 
1 Science Square Newcastle upon Tyne, NE4 5TG\\
Tel: (+44) 191 208 4145 \hskip5em 
Email: m.panagiotopoulou2@newcastle.ac.uk
\item $^{**}$ Yujiang Wang (ORCID ID: 0000-0002-4847-6273)\\
Urban Sciences Building, 
1 Science Square Newcastle upon Tyne, NE4 5TG\\
Tel: (+44) 191 208 4141  \hskip5em
Email: Yujiang.Wang@newcastle.ac.uk
\end{itemize}

\begin{center}
Gabrielle.Schroeder$^a$;
Jessica.Blickwedel$^a$; 
Fahmidaamin.Chowdhury$^b$;
Alison.Cronie$^c$; 
B.Diehl$^d$; 
J.Duncan$^d$;
Jennifer.Falconer$^c$; 
Ryan.Faulder2$^b$;
Veronica.Leach$^c$;
Shona.Livingstone$^c$;
Rhys.Thomas$^a$;
Peter.Taylor$^a$

$^a$ @newcastle.ac.uk
$^b$ @nhs.net
$^c$ @ggc.scot.nhs.uk
$^d$ @ucl.ac.uk
\end{center}

\noindent{
We confirm that we have read the Journal’s position on issues involved in ethical publication and affirm that this report is consistent with those guidelines. \\
None of the authors has any conflict of interest to disclose.}

\newpage
\begin{abstract}

\textbf{Background}
Seizure severity can change from one seizure to the next within individual people with epilepsy. It is unclear if and how seizure severity is modulated over longer timescales. Characterising seizure severity variability over time could lead to tailored treatments. In this study, we test if continuously-recorded interictal intracranial EEG (iEEG) features encapsulate signatures of such modulations. 

\textbf{Methods}
We analysed 20 subjects with iEEG recordings of at least one day. We identified cycles on timescales of hours to days embedded in long-term iEEG band power and associated them with seizure severity, which we approximated using seizure duration. In order to quantify these associations, we created linear-circular statistical models of seizure duration that incorporated different band power cycles within each subject.

\textbf{Findings}
In most subjects, seizure duration was weakly to moderately correlated with individual band power cycles. Combinations of multiple band power cycles significantly explained most of the variability in seizure duration. Specifically, we found $70\%$ of the models had a higher than $60\%$ adjusted $R^2$ across all subjects. From these models, around $80\%$ were deemed to be above chance-level (p-value $\leq$ 0.05) based on permutation tests. Models included cycles of ultradian, circadian and slower timescales in a subject-specific manner.

\textbf{Interpretation}
These results suggest that seizure severity, as measured by seizure duration, may be modulated over timescales of minutes to days by subject-specific cycles in interictal iEEG signal properties. These cycles likely serve as markers of seizure modulating processes. Future work can investigate biological drivers of these detected fluctuations and may inform novel treatment strategies that minimise seizure severity. 

\end{abstract}

\newpage

\section{Introduction}

Seizure severity plays an important role in evaluating therapies for people with epilepsy by identifying which treatments reduce severity. Importantly as the severity of seizures naturally fluctuates over time \citep{lambertsPostictalGeneralizedEEG2013, jobstSecondarilyGeneralizedSeizures2001, pengPostictalGeneralizedEEG2017, gascoigneLibraryQuantitativeMarkers2023}, characterising or forecasting severity in real-time could improve treatment protocols or open new treatment avenues. 
Although multiple tools have been developed to retrospectively quantify seizure severity in individual people with epilepsy \citep{cramerQuantitativeAssessmentSeizure2001, todorovaSEIZURESEVERITYALTERNATIVE2013, bakerDevelopmentSeizureSeverity1991, bakerLiverpoolSeizureSeverity1998, duncanChalfontSeizureSeverity1991, gascoigneLibraryQuantitativeMarkers2023, pattnaikQuantitativeToolSeizure2022a}, there is no established way to forecast seizure severity. 
%Combining this with real-time forecasting of the seizure timing can provide more effective treatment approaches and improve the quality of life with people with epilepsy.

Forecasting seizure severity is challenging since seizure properties change over time within individual people with epilepsy \citep{karolyCyclesEpilepsy2021, schroederSeizurePathwaysChange2020, panagiotopoulouFluctuationsEEGBand2022}. For example, onset locations \citep{gliskeVariabilityLocationHigh2018}, propagation patterns \citep{karthickPredictionSecondaryGeneralization2018}, network evolutions \citep{schroederSeizurePathwaysChange2020, panagiotopoulouFluctuationsEEGBand2022, mitsisFunctionalBrainNetworks2020}, durations \citep{cookHumanFocalSeizures2016, schroederMultipleMechanismsShape2022}, onset times \citep{karolyCyclesEpilepsy2021, baudMultidayRhythmsModulate2018}, and patterns of electrographic epileptiform activity \citep{ilyasIctalHighfrequencyActivity2022} can differ from one seizure to the next. Importantly, seizures with more severe symptoms, such as focal to bilateral tonic clonic (FTBTC) seizures, are more likely to occur at certain times during sleep/wake or day/night cycles \citep{bazilEffectsSleepSleep1997, loddenkemperCircadianPatternsPediatric2011, sinhaSeizuresPatientsRefractory2006, lambertsPostictalGeneralizedEEG2013, jobstSecondarilyGeneralizedSeizures2001, pengPostictalGeneralizedEEG2017}. These findings suggest that seizure properties, including severity, are modulated over short (minutes, hours, and days) and long (weeks, months, and years) timescales \citep{karolyCyclesEpilepsy2021, schroederSeizurePathwaysChange2020, baudMultidayRhythmsModulate2018}. Previous studies have not systematically investigated potential cyclical modulators of seizure severity.  

Cycles in interictal activity could be a biomarker of severity-modulating cycles. Interictal iEEG markers showed prominent cyclical patterns over circadian \citep{panagiotopoulouFluctuationsEEGBand2022, mitsisFunctionalBrainNetworks2020, karolyInterictalSpikesEpileptic2016, karolyCircadianProfileEpilepsy2017, karolyCircadianCircaseptanRhythms2018, spencerCircadianUltradianPatterns2016, karolyCyclesEpilepsy2021} and multi-day \citep{baudMultidayRhythmsModulate2018, baudEndogenousMultidienRhythm2019, karolyInterictalSpikesEpileptic2016, karolyCyclesEpilepsy2021} timescales that appear to influence seizure properties, such as seizure evolution \citep{panagiotopoulouFluctuationsEEGBand2022, schroederSeizurePathwaysChange2020}. It is likely that interictal EEG features and seizures are modulated by common biological factors that cyclically fluctuate over different timescales. As such, cycles of interictal features are a potential biomarker for seizure properties. 

Here we investigate whether seizure severity can be predicted based on interictal features. We use seizure duration as a proxy for seizure severity due to association of seizure duration with clinical seizure types and severity symptoms \citep{dobesbergerDurationFocalComplex2015, jenssenHowLongMost2006, afraDurationComplexPartial2008, kimSeizureDurationDetermined2011, kaufmannWhoSeizesLongest2020, ferastraoaruTerminationSeizureClusters2016}. We analysed changes in seizure duration, and their association with EEG band power fluctuation cycles within subjects. We first examined the association between seizure duration and each band power cycle using correlation metrics. We then assessed the relationship between seizure duration and a combination of band power cycles using a multiple regression framework. Our results demonstrate the relationship between interictal iEEG band power cycles and seizure duration which could provide new opportunities for forecasting seizure severity in the future.

\section{Methods}

\subsection{Patient Cohort and Data Acquisition\label{methods:patientSelec}}

For this study, we analysed iEEG data collected during the presurgical evaluation from 20 adult subjects with refractory focal epilepsy. We used subjects with at least 15 annotated seizures. Data were obtained from the University College London Hospital (UCLH) (13 subjects), the NHS Greater Glasgow and Clyde center (GGC) (three subjects), as well as the Sleep-Wake-Epilepsy-Center (SWEC) at the University Hospital of Bern, Department of Neurology (four subjects) (available at~\href{http://ieeg-swez.ethz.ch}{http://ieeg-swez.ethz.ch}) \citep{burrelloLaelapsEnergyEfficientSeizure2019}. For all subjects, seizures were annotated by clinicians, independently of this study. Seizure durations were measured in seconds, and for analysis converted into log seizure duration using the natural logarithm \citep{schroederMultipleMechanismsShape2022, cookHumanFocalSeizures2016}. Demographic details are given in Supplementary Table \ref{suppl:table_demographic}.

\subsection{iEEG preprocessing and band power computation\label{subsec:preprocessing}}

The iEEG signals from the SWEC cohort were provided in already preprocessed form. Briefly, signals were median-referenced and band-pass filtered from $0.5 - 120$~Hz using a $4^{\text{th}}$ order Butterworth filter (forward and backward). Channels with artifacts were also identified and excluded by the same epileptologist. These steps were all conducted independently of this study and resulted in the publicly available data and annotations \citep{burrelloLaelapsEnergyEfficientSeizure2019}.

For the UCLH and GGC cohorts, we applied similar pre-processing steps before computing the iEEG features of interest. We divided each subject's long-term iEEG data into 30~s non-overlapping time segments. All channels in each time segment were re-referenced to a common average reference. Within each time segment channels that appeared to have outlier amplitude ranges were denoted as noisy and disregarded from the common average calculation.
To remove power line noise, each time segment was notch filtered at $50$~Hz and $100$~Hz. Finally, segments were band-pass filtered from $0.5-120$~Hz (UCLH cohort) or $0.5-110$~Hz (GGC Cohort) using a $4^{\text{th}}$ order zero-phase Butterworth filter (second order forward and backward filter applied). 

For each subject within the SWEC, UCLH, and GGC cohorts, we processed their entire iEEG recording. We extracted iEEG band power from 30~s non-overlapping iEEG segments in five main frequency bands ($\delta:~1-4$~Hz, $\theta:~4-8$ Hz, $\alpha:~8-13$~Hz, $\beta:~13-30$~Hz and $\gamma:~30-80$~Hz) based on Welch's method with 3~s non-overlapping windows.
Missing data were tolerated if one contiguous at least 15~s long segment was available. If missing data were present for one time segment, we applied all processing steps and computed the band power to the non-missing part of this segment.

For each frequency band, extracted continuous band power values were log transformed (natural logarithm), standardised and summarised across all channels by their median. This step resulted in a matrix $A$ for each subject with five rows (number of frequency bands) and number of columns equal to the total number of 30~s time segments in the entire recording. Missing data in $A$ were imputed (see Supplementary section \ref{suppl:ImputeMissing}) to enable extraction of band power cycles on different timescales in the data. Imputed data were not used for later analysis as seizure data were not available in missing periods.

\begin{figure}[H]
    \vspace{-0.9cm}
    \centering
    \includegraphics[scale = 0.9]{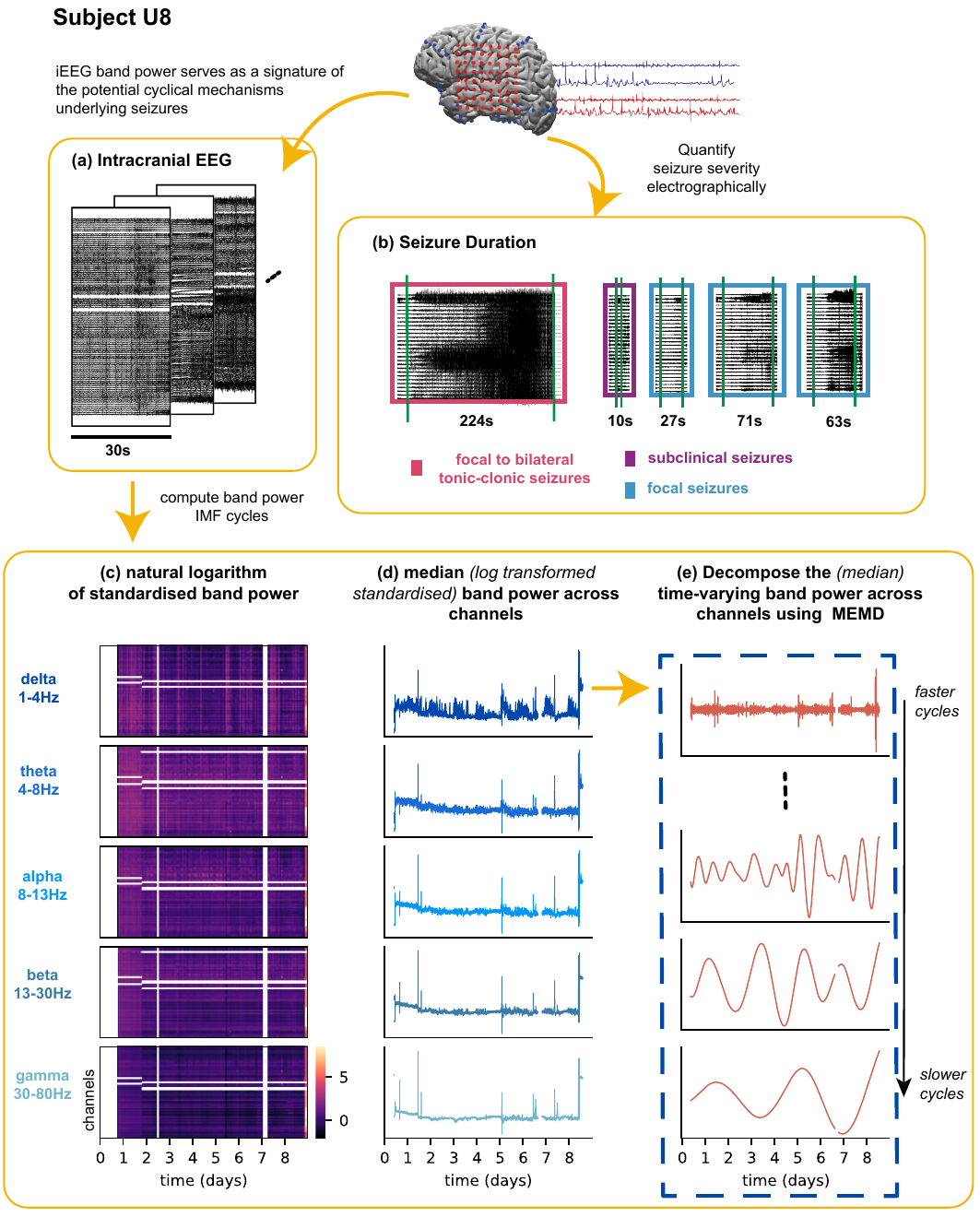}
       \caption{\textbf{Quantifying band power cycles and seizure duration}. (a) Traces of iEEG recordings across channels. Each one corresponds to 30~s of recording. (b) Traces of iEEG recordings for five example seizures with their duration displayed. More severe seizures, such as focal to bilateral tonic-clonic seizures tend to last longer, in comparison with less severe seizures, such as focal or subclinical seizures. (c) Heatmaps of the natural logarithm of (standardised) band power values across channels for each one of the five main frequency bands used in the analysis. (d) Line plots of the median (log transformed standardised) band power across channels against time. Each line plot depicts one frequency band and is coloured accordingly. (e) Visualisation of some example cycles of the delta frequency median band power as extracted using Multivariate Mode Decomposition (MEMD).}
    \label{fig:pipeline}
\end{figure}

\subsection{Delineating iEEG band power cycles using MEMD  \label{subsec:methods_cycles_MEMD}}

To extract oscillatory modes embedded in band power, we applied a signal decomposition method called Empirical Mode Decomposition (EMD)~\citep{huangEmpiricalModeDecomposition1998, huangConfidenceLimitEmpirical2003} that does not require any assumption about stationarity or signal outputs from a linear process. 
EMD captures a finite number, $M$ of narrow-band modes, known as intrinsic mode functions (IMFs), based on the local minima and maxima of the signal. Here the input signal is represented by $Y(t)=\sum_{i=1}^M \text{IMF}_i(t) + r(t)$, where $r(t)$ is the residue signal~\citep{huangEmpiricalModeDecomposition1998}. 
The IMFs are limited to a narrow-band frequency and have a local mean of zero, therefore satisfying the properties needed for Hilbert-transform. Hilbert analysis can be applied to each IMF yielded instantaneous phase (see Supplementary section \ref{suppl:HilbertSpectrum_theory}). As our input matrix $A$ was multidimensional, we applied a multivariate version of EMD, which results in multidimensional IMFs that have the same five dimensions as $A$ (see Supplementary section \ref{suppl:MEMD_theory}). Following the same logic as in the univariate case of EMD (MEMD), each mutlidimensional IMF will reflect a frequency-range related mode across all dimensions (frequency bands). 

Cycles refer to oscillations with a very narrow band frequency representation, whereas fluctuations indicate signals with more variability in their frequency. IMFs are usually narrow-band signals, but due to challenges in the signal, mode mixing may occur in some instances and multiple frequencies may be present. Nevertheless, we will use the term ``cycle'' refer to the IMF obtained from MEMD in following, to be more consistent with previous literature investigating long-term phenomena in interictal iEEG.

\subsection{Statistical analyses}

To test if any band power cycle phase (circular variable) is correlated with seizure duration (linear variable), we first performed Mardia's linear-circular correlation \citep{mardiaLinearCircularCorrelationCoefficients1976}. This results in a correlation coefficient bounded between zero (no relationship) and one (perfect relationship). P-values are calculated from randomisation tests (see Supplementary section \ref{suppl:randomisationTestCorr}). FDR correction (see Supplementary section \ref{suppl:FDR}) was applied to all p-values across all individuals and tests; the significance level was set to 5\%.

To test if a combination of band power cycles are able to explain seizure duration, we applied Linear-circular regression (see Supplementary section \ref{suppl:CircRegression}). We applied a model selection process (see Supplementary section \ref{suppl:formingModels}). Model performance was assessed using root-mean-squared-error (RMSE) and adjusted $R^2$ as model performance metrics. A range of models were tested in the selection process, including all band power cycles (all IMFs) in the delta, theta, alpha, beta and gamma band individually, as well as the strongest cycles embedded in the signals (cycles with very high power across all dimensions (frequency bands) - we term the resultant models as ``peak'' models.
Finally, we tested if the variable selection and regression process for the selected model could have yielded a high performance by chance by running a permutation test (see Supplementary section \ref{suppl:permutation_performance} for details). The significance level was set to 5\%.

\section{Code and data availability}

Code and data will be made available upon acceptance of the paper.

%%%%%%%%%%%%%%%%%%%% RESULTS %%%%%%%%%%%%%%%%%%%%%
\section{Results}
\subsection{Association of seizure duration and individual band power cycles}

\begin{figure}[h!]
    %\hspace{-0.2cm}
    \centering
    \includegraphics[scale = 0.9]{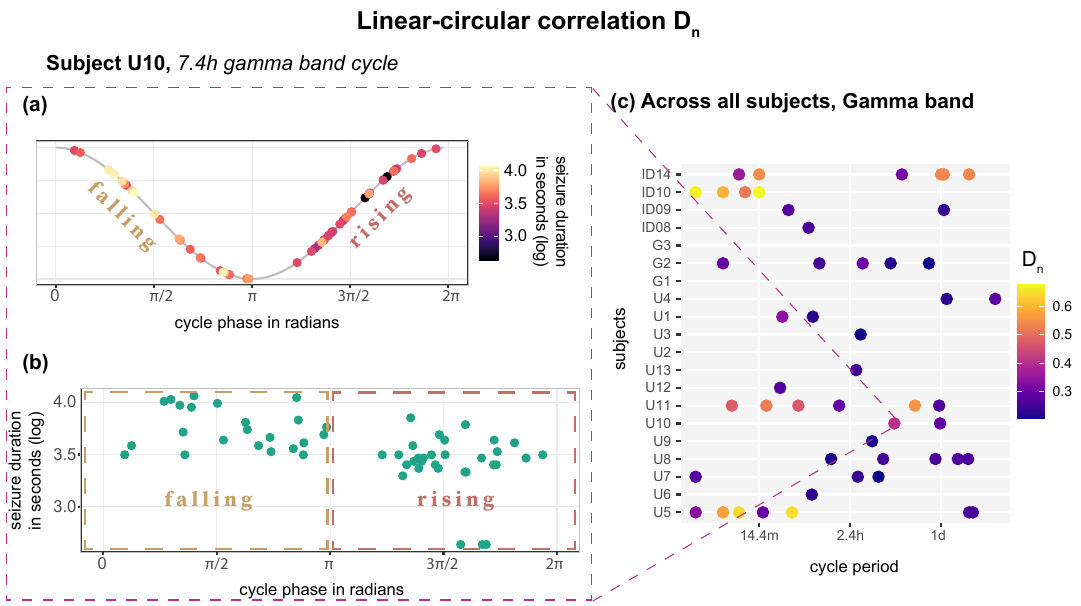}
       \caption{\textbf{Example associations of seizure duration with phases of band power cycles}. (a) Representation of the phase of the 7.4h gamma band power cycle  $[0, 2\pi]$. The curve illustrates the falling and rising phase of the band power cycle. Each point represents one seizure that occurred at the corresponding phase of the band power cycle and are coloured by seizure duration in seconds (log). (b) Alternative, scatter plot representation of the same data as in (a): seizure duration shown against phases of the band power cycle. (c) Dot plot of the Mardia's rank correlation between the seizure duration and the phases of the band power cycles across all subjects. Weak associations ($D_{n} \leq 0.2$) are not shown for clarity of visualisation.}
    \label{fig:Mardias_corr}
\end{figure}

For each subject, we first determined any association between the seizure duration and the phase of each band power cycle at which the seizure occurred using the rank linear-circular correlation $D_{n}$ \citep{mardiaLinearCircularCorrelationCoefficients1976}. In example patient U10, Fig. \ref{fig:Mardias_corr}a and b illustrate a moderate correlation ($D_{n} = 0.40$, $p = 0.0093$) between the seizure duration and a gamma band power cycle (7.4h characteristic cycle period). Longer seizures tended to occur during the falling phase of the cycle (Fig.~\ref{fig:Mardias_corr}a and b). %However, after applying FDR for multiple comparisons across all associations between seizure duration and phases of band power cycles for all subjects, this association was not significant (q-value $= 0.175$) for a $5\%$ significance level. 
% (c) Table of information related to the association of the seizure duration with the example band power cycle. This table contains the non-parametric circular-linear correlation (Mardia's rank correlation) $D_{n}$, the significance (p-value) of this correlation as obtained from a permutation test, the significance of the correlation after FDR correction for multiple comparisons 
%This is also evident from $95\%$ confidence interval obtained using bootstrap percentile confidence interval (see Section \ref{subsubsec:BootCI}) where a high variation of our estimate is apparent.
Fig.~\ref{fig:Mardias_corr}c shows the correlations between the seizure duration and the phases of gamma band power cycles across subjects. Overall, there are some weak to moderate correlations with some stronger ones appearing for two subjects (ID10 and U5). 

Across subjects, we mostly found weak to moderate correlations between the seizure duration and the phases of individual band power cycles for every frequency band (see Supplementary Fig.~\ref{fig:Mardias_corr_all_freq_bands}). %As illustrated in Fig. \ref{fig:Mardias_corr}c, only five subjects appeared to have stronger correlations with the phases of one or more band power cycles (ID14, ID10, 1200, 1182, 1005).

\newpage

\subsection{Seizure duration is modulated by cycles on different timescales}

We further investigated whether seizure duration is more strongly associated with a combination of two or more band power cycles, as individual cycles only showed weak to moderate associations (Fig.~\ref{fig:Mardias_corr}c and Suppl. Fig.~\ref{fig:Mardias_corr_all_freq_bands}). To uncover such associations, we applied a circular-linear model selection and regression (see Supplementary section \ref{suppl:CircRegression}). Fig.~\ref{fig:models_one_example} illustrates the selected model for an example subject, U4. This model had the best fit on the data with an adjusted $R^2 \text{ of } 82\%$ (Fig. \ref{fig:models_one_example}a). The performance of this model can be seen in Fig. \ref{fig:models_one_example}b, where the predicted seizure durations based on band power cycles were close to the actual seizure duration values. Fig. \ref{fig:models_one_example}c illustrates how much each band power cycle contributed to the seizure duration predictions based on the final model.
% For visualisation, in Fig. \ref{fig:models_one_example}d, we additionally show the raw iEEG of two example band power cycles, IMF 9 and 11. All the variables included in the model over the full spectrum of all different timescales and frequency bands are shown in Fig. \ref{fig:models_one_example}c. As can be seen in Fig. \ref{fig:models_one_example}c, six faster cycles (days/cycle $<$ 1 day) were included, as well as four slower cycles of timescales ranging from $\sim$ 1.2 days to $\sim$ 4 days.

\begin{figure}[h!]
    % \hspace{-1.1cm}
    \centering
    \includegraphics[scale = 0.9]{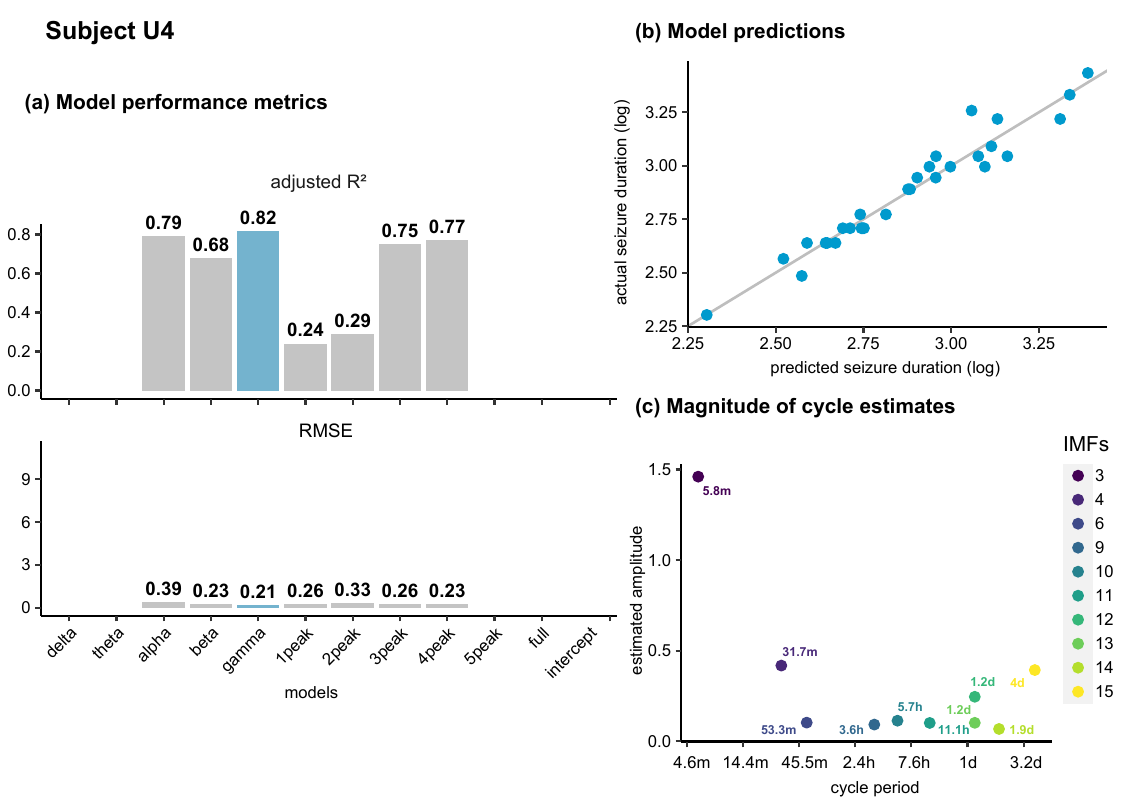}
       \caption{\textbf{Illustration of the final selected model for one example subject}. (a) Bar plots representing the performance of the different models evaluated for this example subject. The top plot shows the $\text{adjusted}~R^2$, while the bottom plot shows the root-mean-squared-error (RMSE) as obtained from the Leave-one-out-cross-validation (LOOCV). The non-grey coloured bars indicate the selected model, while all the other models are displayed with grey colour. Performance metrics for models that were not valid are not shown. (b) Scatter plots depicting the estimated quantities obtained from the model against the actual seizure duration (natural logarithm) in this patient. (c) Band power cycle importance as expressed by its estimated amplitude/magnitude (see Supplementary section \ref{suppl:CircRegression}) for the corresponding cycle included in the model plotted against their corresponding cycle period (days/cycle).}
    \label{fig:models_one_example}
\end{figure}

\begin{figure}[h!]
    % \hspace{-0.5cm}
    \centering
    \includegraphics[scale = 0.9]{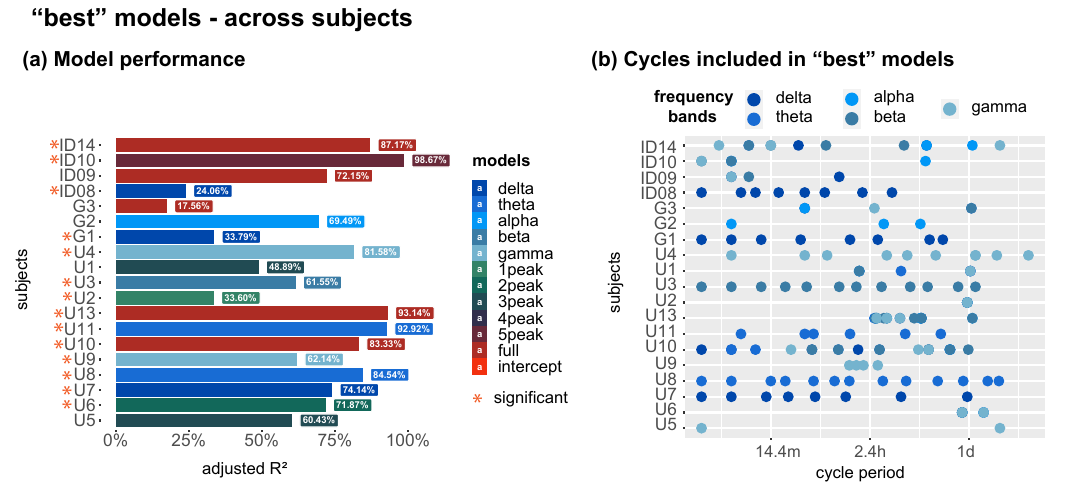}
       \caption{\textbf{Selected models across subjects}. (a) Barplot of the adjusted $R^2$ as obtained from the selected model for each subject. The significance is indicated with a red asterisk as determined by a permutation test for the adjusted $R^2$ (see Suppl. Fig.~\ref{fig:perm_A}).
      (b) Dot plot illustrating the cycle period of the variables selected in the final model for each subject. Colours depict the frequency band associated with these band power cycles. %Subject 1211 is not displayed, as for this subject the "intercept" model (model containing just the intercept) performed best, which means that no band power cycle have any additional contribution on  explaining the variability of seizure duration rather than the sampling mean.
      }
    \label{fig:summary_models}
\end{figure}

Across subjects, we found $70\%$ of the subject-specific models had an adjusted $R^2$ over $60\%$ (Fig.~\ref{fig:summary_models}a). From these models, around $80\%$ were deemed to be above chance-level (p-value $<=$ 0.05) based on permutation tests (see Fig.~\ref{fig:summary_models}a, Supplementary Fig.~\ref{fig:perm_A}).
The intercept model performed better compared to all the models in only one subject (U12), indicating that this subject's variability in seizure duration was not well-predicted by band power cycles. 
Overall, these models could explain more than $60\%$ of the variability captured in seizure duration for those subjects. 

Different band power cycles of various timescales contributed in seizure duration variability in the selected models (Fig.~\ref{fig:summary_models}b). Timescales less than 1 day contributed to the seizure duration diversity in all 19 subjects. For subjects U9 and ID10 the circadian cycle was not apparent. Thus, circadian band power cycle was detected and included in the models in 17 subjects out of 19. In 11 out of these 17 subjects the circadian rhythm contributed in the selected models. For four subjects, slower cycles with cycle period 2 or 4 days were included in the modeling process (subject (period of slower cycle in days): ID14 (2 days), U4 (4 days), U8 (2 days) and U5 (2.3 days)) (see Suppl. section \ref{suppl:choice_of_cycles}), those appeared to have cycles with periods longer than one day in the selected models.

%%%%%%%%%%%%%%%%%%%% DISCUSSION %%%%%%%%%%%%%%%%%%%%%

\section{Discussion}

A combination of cycles in interictal band power can explain the variability in seizure severity, as measured by seizure duration, within individual subjects. We found that a range of timescales provided explanatory power, including faster cycles (cycle period $<$ 1 day), circadian cycles (cycle period $=$ 1 day) and some slower cycles (cycle period $>$ 1 day). It is likely that a combination of oscillatory mechanisms, each with distinct cycle period, shapes how seizure severity changes over time. The incorporation of these findings into therapeutic interventions, such as chronotherapy \citep{carneyChronotherapyTreatmentEpilepsy2014, ramgopalChronopharmacologyAntiConvulsiveTherapy2013} may mitigate seizure severity.
    
Although a combination of different band power cycles could explain seizure duration, this relationship should not be interpreted as a causal effect. Each band power cycle is most likely a reflection of an underlying modulatory process. It is not surprising that a combination of different timescales predicts seizure duration, as multiple mechanisms \citep{panagiotopoulouFluctuationsEEGBand2022, schroederSeizurePathwaysChange2020} appear to modulate features such as seizure occurrence \citep{karolyCyclesEpilepsy2021, bernardSeizuresRightTime2021} and seizure duration \citep{schroederMultipleMechanismsShape2022}. 

The nature of these modulatory processes is still elusive, although their timescales may provide some insights. Circadian cycles might be due to molecular oscillations \citep{bernardCircadianMultidienMolecular2021} or may be linked to the sleep-wake cycle \citep{jinEpilepsyItsInteraction2020a}. Shorter timescales might be related to blood concentrations of metabolites, as glucose and ketone bodies fluctuate in timescales ranging from minutes to hours \citep{verbeekHourlyAnalysisCerebrospinal2016, simeoneKetoneBodiesMediate2018}. These metabolic substances have also been linked to neural hyperexcitability \citep{simeoneKetoneBodiesMediate2018}.
Thus, cycles in iEEG markers such as band power could serve as biomarkers for oscillatory mechanisms that influence brain dynamics; these mechanisms might in turn modulate the initiation, severity, and/or termination of seizures.
% I was thinking of finding clues regarding potential mechanisms that might affect seizure severity directly or seizure duration etc.....
% cycles review Karoly, 10 misconceptions Karoly, Leguia, Chronotypes
% https://pubmed.ncbi.nlm.nih.gov/30776337/
% https://www.ncbi.nlm.nih.gov/pmc/articles/PMC6017683/

Although we found many associations between band power cycles and seizure duration, we were unable to explain duration variability in a small number of individuals. Incorporation of longer cycles (than the ones captured in our data) into the models might contribute to the unexplained variability of seizure duration. However, the lack of ultra long-term iEEG recordings (several weeks or months) limited our analysis to multiday cycles up to four days (see Supplementary figures \ref{fig:char_ampl_freq1} and \ref{fig:char_ampl_freq2}). In previous studies, cycles over a span of 5.5-33 days \citep{baudMultidayRhythmsModulate2018} have been associated with seizure occurrence. Additionally, multi-month cycles have been associated with changes in features of seizure evolutions \citep{schroederChronicIEEGRecordings2022}. Another explanation is some level of random variability (stochasticity) on seizure duration~\citep{silvaEpilepsiesDynamicalDiseases2003, suffczynskiDynamicsEpilepticPhenomena2006}; this stochastic element cannot be predicted. Lastly, other non-rhythmic factors could contribute to variable seizure duration. For example, environmental and physiological factors \citep{payneIdentifyingSeizureRisk2021} have been shown to affect seizure occurrence. Anti-seizure drug withdrawal could be another contributing factor \citep{kirbyDrugWithdrawalEpilepsy2020, spencerIctalEffectsAnticonvulsant1981, zhouInfluenceIctalSeizure2002, duyAntiepilepticDrugWithdrawal2020}. These external and internal factors might also impact other seizure characteristics, such as seizure duration and severity. A model that quantifies these factors may achieve better predictive performance of the seizure duration.

Limitations: In this study, we analysed a broad spectrum of spectral properties of brain activity represented by a time-varying band power. We focused on the temporal changes by using the median band power across channels, providing an overview of the possible time-scales of modulatory cycles. However, variability of seizure severity might be linked to specific brain regions where band power cycles have a unique signature. For example, more severe seizures, such as FTBTC seizures have been linked to subcortical areas, such as the thalamus \citep{caciagliThalamusFocalBilateral2020}.
%You: Not sure whether the iEEG electrodes placed in the thalamus as well. Amygdala and hippocampus are deeper sources that measured with iEEG
Future work may therefore benefit from including spatial information from iEEG or other modalities.
%Cortical areas, such as frontal and temporal lobe have been associated with the phenomenon of seizure clustering \citep{ferastraoaruTerminationSeizureClusters2016}, yielding to more severe seizure manifestations, as seizure clusters are associated with an increased risk of prolonged seizures, status epilepticus \citep{hautSeizureClusteringRisks2005}, and/or postictal morbidity, such as psychosis \citep{roseOccurrenceSeizureClusters2003}. Interestingly, seizure clustering reveals a predisposition on distinct circadian cycles/signatures between frontal and temporal lobe seizures, as it occurs at different times between the two sites \citep{pavlovaDiurnalPatternSeizures2012}. Spatial patterns of various iEEG features, such as iEEG band power \citep{panagiotopoulouFluctuationsEEGBand2022}, high frequency activity \citep{gliskeVariabilityLocationHigh2018}, and interictal spike rate \citep{conradSpatialDistributionInterictal2020} change over time. Interestingly, band power changes have been associated with seizure spread in areas outside of the seizure focus, suggesting a widespread activity might manifest more severe seizures, such as secondarily generalized ones \citep{naftulinIctalPreictalPower2018}. Future work can disentangle the spatio-temporal representation of cycles embedded in interictal iEEG features in order to identify oscillatory activity of specific brain regions that could explain variability in seizure severity. 

Another limitation is that we used seizure duration as a proxy for seizure severity. Even though seizure duration has been linked to seizure types \citep{kaufmannWhoSeizesLongest2020}, postictal activity \citep{paynePostictalSuppressionSeizure2018}, and SUDEP \citep{ochoa-urreaSeizureClustersSeizure2021}, it does not encapsulate all aspects of seizure severity. Various other aspects of seizure severity could be quantified electrographically, such as the strength and extent of spread of ictal activity, as well as the duration of post-ictal suppression \citep{gascoigneLibraryQuantitativeMarkers2023}. Additionally, other physiological markers of seizure severity have been used in previous studies \citep{gascoigneLibraryQuantitativeMarkers2023}. As with seizure duration, individual markers of seizure severity can be associated with band power cycles using our approach (Supplementary Fig. \ref{fig:top_delta_summary} shows one example). However, combining different severity markers may lead to a more representative overall measure of severity for better evaluating its temporal cyclical patterns. Identifying potential endogenous mechanisms (physiological and cognitive) or/and external factors that fluctuate on similar periodicities might unravel the physiological drivers of variable seizure severity. 
     
    %  \item additional content for why we used seizure duration as proxy for seizure severity - maybe condense with previous paragraph\\
    %  Seizure duration has been linked to seizure types \citep{dobesbergerDurationFocalComplex2015, jenssenHowLongMost2006, afraDurationComplexPartial2008, kimSeizureDurationDetermined2011}: Seizure types with more severe symptoms tend to last longer. For example seizures that start in one area on one side of the brain (focal) and spread across the brain, recruiting additional brain areas (bilateral tonic-clonic) tend to have longer duration compared to those that do not spread further \citep{dobesbergerDurationFocalComplex2015, kaufmannWhoSeizesLongest2020, ferastraoaruTerminationSeizureClusters2016}. Longer seizures have also been associated with extensive periods of postictal suppression (an undisputed marker of seizure burden and severity) compared to short seizures \citep{paynePostictalSuppressionSeizure2018}. Thus, we will use seizure duration as an intermediate proxy of seizure severity. 
    
Finally, we were limited to a relatively short recording period of a few days, possibly missing longer cycles. The number of seizures in each patient was also relatively low, and we therefore did not attempt testing the performance of a possible predictor. Although we could show statistically that our association was above chance-level, future work should test whether prospective prediction is feasible.

In conclusion, we have observed that seizure severity can be explained by a range of cycles on different timescales captured in time-varying iEEG band power. We conclude that prospective predictions may be possible in future with longer recordings, possibly including other modalities. Our results provide evidence for multiple modulatory cycles on different timescales that impact seizures and their severity, which future work may investigate further. Ultimately, a better understanding of the seizure modulating processes will enable the development of novel treatment strategies that could minimise seizure severity and therefore the clinical impact of seizures.

%Identifying the patterns that seizures shift from more severe seizures to less ones can lead to the actual pathophysiological mechanisms; mild seizures might be modulated by different mechanisms compared to the more severe ones, but a combination of fluctuations of different timescales can explain the full spectrum of the seizure severity. 

%%%%%%%%%%%%%%%%%%%% ACKNOWLEDGEMENTS %%%%%%%%%%%%%%%%%%%%%
\section{Acknowledgements}

We thank members of the Computational Neurology, Neuroscience \& Psychiatry Lab (www.cnnp-lab.com) for discussions on the analysis and manuscript; P.N.T. and Y.W. are both supported by UKRI Future Leaders Fellowships (MR/T04294X/1, MR/V026569/1). MP is supported by the Centre for Doctoral Training in Cloud Computing for Big Data (EP/L015358/1). JSD, JdT are supported by the NIHR UCLH/UCL Biomedical Research Centre. B.D. receives support from the NIH National Institute of Neurological Disorders and Stroke U01-NS090407 (Center for SUDEP Research) and Epilepsy Research UK. 

\newpage
\bibliography{THESIS}

\begin{thebibliography}{71}
\providecommand{\natexlab}[1]{#1}
\providecommand{\url}[1]{\texttt{#1}}
\expandafter\ifx\csname urlstyle\endcsname\relax
  \providecommand{\doi}[1]{doi: #1}\else
  \providecommand{\doi}{doi: \begingroup \urlstyle{rm}\Url}\fi

\bibitem[Afra et~al.(2008)Afra, Jouny, and
  Bergey]{afraDurationComplexPartial2008}
Pegah Afra, Christophe~C. Jouny, and Gregory~K. Bergey.
\newblock Duration of complex partial seizures: {{An}} intracranial {{EEG}}
  study.
\newblock \emph{Epilepsia}, 49\penalty0 (4):\penalty0 677--684, 2008.
\newblock ISSN 1528-1167.
\newblock \doi{10.1111/j.1528-1167.2007.01420.x}.
\newblock URL
  \url{http://onlinelibrary.wiley.com/doi/abs/10.1111/j.1528-1167.2007.01420.x}.

\bibitem[Arlot and Celisse(2010)]{arlotSurveyCrossvalidationProcedures2010}
Sylvain Arlot and Alain Celisse.
\newblock A survey of cross-validation procedures for model selection.
\newblock \emph{Statistics Surveys}, 4\penalty0 (none), January 2010.
\newblock ISSN 1935-7516.
\newblock \doi{10.1214/09-SS054}.
\newblock URL
  \url{https://projecteuclid.org/journals/statistics-surveys/volume-4/issue-none/A-survey-of-cross-validation-procedures-for-model-selection/10.1214/09-SS054.full}.

\bibitem[Baker et~al.(1991)Baker, Smith, Dewey, Morrow, Crawford, and
  Chadwick]{bakerDevelopmentSeizureSeverity1991}
Gus~A Baker, David~F Smith, Michael Dewey, James Morrow, Pamela~M Crawford, and
  David~W Chadwick.
\newblock The development of a seizure severity scale as an outcome measure in
  epilepsy.
\newblock \emph{Epilepsy Research}, 8\penalty0 (3):\penalty0 245--251, April
  1991.
\newblock ISSN 09201211.
\newblock \doi{10.1016/0920-1211(91)90071-M}.
\newblock URL
  \url{https://linkinghub.elsevier.com/retrieve/pii/092012119190071M}.

\bibitem[Baker et~al.(1998)Baker, Smith, Jacoby, Hayes, and
  Chadwick]{bakerLiverpoolSeizureSeverity1998}
Gus~A. Baker, David~F. Smith, Ann Jacoby, Julie~A. Hayes, and David~W.
  Chadwick.
\newblock Liverpool {{Seizure Severity Scale}} revisited.
\newblock \emph{Seizure}, 7\penalty0 (3):\penalty0 201--205, June 1998.
\newblock ISSN 10591311.
\newblock \doi{10.1016/S1059-1311(98)80036-8}.
\newblock URL
  \url{https://linkinghub.elsevier.com/retrieve/pii/S1059131198800368}.

\bibitem[Baud et~al.(2018)Baud, Kleen, Mirro, Andrechak, {King-Stephens},
  Chang, and Rao]{baudMultidayRhythmsModulate2018}
Maxime~O. Baud, Jonathan~K. Kleen, Emily~A. Mirro, Jason~C. Andrechak, David
  {King-Stephens}, Edward~F. Chang, and Vikram~R. Rao.
\newblock Multi-day rhythms modulate seizure risk in epilepsy.
\newblock \emph{Nature Communications}, 9\penalty0 (1):\penalty0 1--10, January
  2018.
\newblock ISSN 2041-1723.
\newblock \doi{10.1038/s41467-017-02577-y}.
\newblock URL \url{https://www.nature.com/articles/s41467-017-02577-y}.

\bibitem[Baud et~al.(2019)Baud, Ghestem, Benoliel, Becker, and
  Bernard]{baudEndogenousMultidienRhythm2019}
Maxime~O. Baud, Antoine Ghestem, Jean-Jacques Benoliel, Christel Becker, and
  Christophe Bernard.
\newblock Endogenous multidien rhythm of epilepsy in rats.
\newblock \emph{Experimental Neurology}, 315:\penalty0 82--87, May 2019.
\newblock ISSN 0014-4886.
\newblock \doi{10.1016/j.expneurol.2019.02.006}.
\newblock URL
  \url{https://www.sciencedirect.com/science/article/pii/S0014488618302838}.

\bibitem[Bazil and Walczak(1997)]{bazilEffectsSleepSleep1997}
Carl~W. Bazil and Thaddeus~S. Walczak.
\newblock Effects of {{Sleep}} and {{Sleep Stage}} on {{Epileptic}} and
  {{Nonepileptic Seizures}}.
\newblock \emph{Epilepsia}, 38\penalty0 (1):\penalty0 56--62, 1997.
\newblock ISSN 1528-1167.
\newblock \doi{10.1111/j.1528-1157.1997.tb01077.x}.
\newblock URL
  \url{https://onlinelibrary.wiley.com/doi/abs/10.1111/j.1528-1157.1997.tb01077.x}.

\bibitem[Benjamini and Hochberg(1995)]{benjaminiControllingFalseDiscovery1995}
Yoav Benjamini and Yosef Hochberg.
\newblock Controlling the {{False Discovery Rate}}: {{A Practical}} and
  {{Powerful Approach}} to {{Multiple Testing}}.
\newblock \emph{Journal of the Royal Statistical Society: Series B
  (Methodological)}, 57\penalty0 (1):\penalty0 289--300, January 1995.
\newblock ISSN 00359246.
\newblock \doi{10.1111/j.2517-6161.1995.tb02031.x}.
\newblock URL
  \url{https://onlinelibrary.wiley.com/doi/10.1111/j.2517-6161.1995.tb02031.x}.

\bibitem[Bernard(2021)]{bernardCircadianMultidienMolecular2021}
Christophe Bernard.
\newblock Circadian/multidien {{Molecular Oscillations}} and {{Rhythmicity}} of
  {{Epilepsy}} ({{MORE}}).
\newblock \emph{Epilepsia}, 62\penalty0 (1):\penalty0 S49--S68, 2021.
\newblock ISSN 1528-1167.
\newblock \doi{10.1111/epi.16716}.
\newblock URL \url{http://onlinelibrary.wiley.com/doi/abs/10.1111/epi.16716}.

\bibitem[Bernard and Nehlig(2021)]{bernardSeizuresRightTime2021}
Christophe Bernard and Astrid Nehlig.
\newblock Seizures: {{About}} the right time to explore their mechanisms.
\newblock \emph{Epilepsia}, 62\penalty0 (S1):\penalty0 S1--S1, 2021.
\newblock ISSN 1528-1167.
\newblock \doi{10.1111/epi.16782}.
\newblock URL \url{http://onlinelibrary.wiley.com/doi/abs/10.1111/epi.16782}.

\bibitem[Breheny and Huang(2015)]{brehenyGroupDescentAlgorithms2015}
Patrick Breheny and Jian Huang.
\newblock Group descent algorithms for nonconvex penalized linear and logistic
  regression models with grouped predictors.
\newblock \emph{Statistics and Computing}, 25\penalty0 (2):\penalty0 173--187,
  March 2015.
\newblock ISSN 1573-1375.
\newblock \doi{10.1007/s11222-013-9424-2}.
\newblock URL \url{https://doi.org/10.1007/s11222-013-9424-2}.

\bibitem[BURMAN(1989)]{burmanComparativeStudyOrdinary1989}
{\relax PRABIR}~BURMAN.
\newblock A comparative study of ordinary cross-validation, v-fold
  cross-validation and the repeated learning-testing methods.
\newblock \emph{Biometrika}, 76\penalty0 (3):\penalty0 503--514, September
  1989.
\newblock ISSN 0006-3444.
\newblock \doi{10.1093/biomet/76.3.503}.
\newblock URL \url{https://doi.org/10.1093/biomet/76.3.503}.

\bibitem[Burrello et~al.(2019)Burrello, Cavigelli, Schindler, Benini, and
  Rahimi]{burrelloLaelapsEnergyEfficientSeizure2019}
Alessio Burrello, Lukas Cavigelli, Kaspar Schindler, Luca Benini, and Abbas
  Rahimi.
\newblock Laelaps: {{An Energy-Efficient Seizure Detection Algorithm}} from
  {{Long-term Human iEEG Recordings}} without {{False Alarms}}.
\newblock In \emph{2019 {{Design}}, {{Automation}} \& {{Test}} in {{Europe
  Conference}} \& {{Exhibition}} ({{DATE}})}, pages 752--757, {Florence,
  Italy}, March 2019. {IEEE}.
\newblock ISBN 978-3-9819263-2-3.
\newblock \doi{10.23919/DATE.2019.8715186}.
\newblock URL \url{https://ieeexplore.ieee.org/document/8715186/}.

\bibitem[Caciagli et~al.(2020)Caciagli, Allen, He, Trimmel, Vos, Centeno,
  Galovic, Sidhu, Thompson, Bassett, Winston, Duncan, Koepp, and
  Sperling]{caciagliThalamusFocalBilateral2020}
Lorenzo Caciagli, Luke~A. Allen, Xiaosong He, Karin Trimmel, Sjoerd~B. Vos,
  Maria Centeno, Marian Galovic, Meneka~K. Sidhu, Pamela~J. Thompson,
  Danielle~S. Bassett, Gavin~P. Winston, John~S. Duncan, Matthias~J. Koepp, and
  Michael~R. Sperling.
\newblock Thalamus and focal to bilateral seizures.
\newblock \emph{Neurology}, 95\penalty0 (17):\penalty0 e2427--e2441, October
  2020.
\newblock ISSN 0028-3878.
\newblock \doi{10.1212/WNL.0000000000010645}.
\newblock URL \url{https://www.ncbi.nlm.nih.gov/pmc/articles/PMC7682917/}.

\bibitem[Carney et~al.(2014)Carney, Stanley, and
  Talathi]{carneyChronotherapyTreatmentEpilepsy2014}
Paul Carney, David Stanley, and Sachin Talathi.
\newblock Chronotherapy in the treatment of epilepsy.
\newblock \emph{ChronoPhysiology and Therapy}, 4:\penalty0 109--123, November
  2014.
\newblock ISSN 2230-2026.
\newblock \doi{10.2147/CPT.S54530}.
\newblock URL
  \url{http://www.dovepress.com/chronotherapy-in-the-treatment-of-epilepsy-peer-reviewed-article-CPT}.

\bibitem[Cook et~al.(2016)Cook, Karoly, Freestone, Himes, Leyde, Berkovic,
  O'Brien, Grayden, and Boston]{cookHumanFocalSeizures2016}
Mark~J. Cook, Philippa~J. Karoly, Dean~R. Freestone, David Himes, Kent Leyde,
  Samuel Berkovic, Terence O'Brien, David~B. Grayden, and Ray Boston.
\newblock Human focal seizures are characterized by populations of fixed
  duration and interval.
\newblock \emph{Epilepsia}, 57\penalty0 (3):\penalty0 359--368, March 2016.
\newblock ISSN 00139580.
\newblock \doi{10.1111/epi.13291}.
\newblock URL \url{https://onlinelibrary.wiley.com/doi/10.1111/epi.13291}.

\bibitem[Cramer and French(2001)]{cramerQuantitativeAssessmentSeizure2001}
Joyce~A. Cramer and Jacqueline French.
\newblock Quantitative {{Assessment}} of {{Seizure Severity}} for {{Clinical
  Trials}}: {{A Review}} of {{Approaches}} to {{Seizure Components}}.
\newblock \emph{Epilepsia}, 42\penalty0 (1):\penalty0 119--129, 2001.
\newblock ISSN 1528-1167.
\newblock \doi{10.1046/j.1528-1157.2001.19400.x}.
\newblock URL
  \url{http://onlinelibrary.wiley.com/doi/abs/10.1046/j.1528-1157.2001.19400.x}.

\bibitem[Dobesberger et~al.(2015)Dobesberger, Risti{\'c}, Walser, Kuchukhidze,
  Unterberger, H{\"o}fler, Amann, and
  Trinka]{dobesbergerDurationFocalComplex2015}
Judith Dobesberger, Aleksandar~J. Risti{\'c}, Gerald Walser, Giorgi
  Kuchukhidze, Iris Unterberger, Julia H{\"o}fler, Edda Amann, and Eugen
  Trinka.
\newblock Duration of focal complex, secondarily generalized tonic\textendash
  clonic, and primarily generalized tonic\textendash clonic seizures
  \textemdash{} {{A}} video-{{EEG}} analysis.
\newblock \emph{Epilepsy \& Behavior}, 49:\penalty0 111--117, August 2015.
\newblock ISSN 1525-5050.
\newblock \doi{10.1016/j.yebeh.2015.03.023}.
\newblock URL
  \url{https://www.sciencedirect.com/science/article/pii/S1525505015001286}.

\bibitem[Duncan and Sander(1991)]{duncanChalfontSeizureSeverity1991}
J~S Duncan and J~W Sander.
\newblock The {{Chalfont Seizure Severity Scale}}.
\newblock \emph{Journal of Neurology, Neurosurgery \& Psychiatry}, 54\penalty0
  (10):\penalty0 873--876, October 1991.
\newblock ISSN 0022-3050.
\newblock \doi{10.1136/jnnp.54.10.873}.
\newblock URL \url{https://jnnp.bmj.com/lookup/doi/10.1136/jnnp.54.10.873}.

\bibitem[Duy et~al.(2020)Duy, Krauss, Crone, Ma, and
  Johnson]{duyAntiepilepticDrugWithdrawal2020}
Phan~Q. Duy, Gregory~L. Krauss, Nathan~E. Crone, Molly Ma, and Emily~L.
  Johnson.
\newblock Antiepileptic drug withdrawal and seizure severity in the epilepsy
  monitoring unit.
\newblock \emph{Epilepsy \& Behavior}, 109:\penalty0 107128, August 2020.
\newblock ISSN 1525-5050.
\newblock \doi{10.1016/j.yebeh.2020.107128}.
\newblock URL
  \url{https://www.sciencedirect.com/science/article/pii/S1525505020303073}.

\bibitem[Ferastraoaru et~al.(2016)Ferastraoaru, {Schulze-Bonhage}, Lipton,
  D{\"u}mpelmann, Legatt, Blumberg, and
  Haut]{ferastraoaruTerminationSeizureClusters2016}
Victor Ferastraoaru, Andreas {Schulze-Bonhage}, Richard~B. Lipton, Matthias
  D{\"u}mpelmann, Alan~D. Legatt, Julie Blumberg, and Sheryl~R. Haut.
\newblock Termination of seizure clusters is related to the duration of focal
  seizures.
\newblock \emph{Epilepsia}, 57\penalty0 (6):\penalty0 889--895, June 2016.
\newblock ISSN 1528-1167.
\newblock \doi{10.1111/epi.13375}.

\bibitem[Fingelkurts and
  Fingelkurts(2001)]{fingelkurtsOperationalArchitectonicsHuman2001}
Andrew~A. Fingelkurts and Alexander~A. Fingelkurts.
\newblock Operational {{Architectonics}} of the {{Human Brain Biopotential
  Field}}: {{Towards Solving}} the {{Mind-Brain Problem}}.
\newblock \emph{Brain and Mind}, 2\penalty0 (3):\penalty0 261--296, December
  2001.
\newblock ISSN 1573-3300.
\newblock \doi{10.1023/A:1014427822738}.
\newblock URL \url{https://doi.org/10.1023/A:1014427822738}.

\bibitem[Gascoigne et~al.(2023)Gascoigne, Waldmann, Schroeder, Panagiotopoulou,
  Blickwedel, Chowdhury, Cronie, Diehl, Duncan, Falconer, Faulder, Guan, Leach,
  Livingstone, Papasavvas, Thomas, Wilson, Taylor, and
  Wang]{gascoigneLibraryQuantitativeMarkers2023}
Sarah~J. Gascoigne, Leonard Waldmann, Gabrielle~M. Schroeder, Mariella
  Panagiotopoulou, Jess Blickwedel, Fahmida Chowdhury, Alison Cronie, Beate
  Diehl, John~S. Duncan, Jennifer Falconer, Ryan Faulder, Yu~Guan, Veronica
  Leach, Shona Livingstone, Christoforos Papasavvas, Rhys~H. Thomas, Kevin
  Wilson, Peter~N. Taylor, and Yujiang Wang.
\newblock A library of quantitative markers of seizure severity.
\newblock \emph{Epilepsia}, page epi.17525, February 2023.
\newblock ISSN 0013-9580, 1528-1167.
\newblock \doi{10.1111/epi.17525}.
\newblock URL \url{https://onlinelibrary.wiley.com/doi/10.1111/epi.17525}.

\bibitem[Gliske et~al.(2018)Gliske, Irwin, Chestek, Hegeman, Brinkmann, Sagher,
  Garton, Worrell, and Stacey]{gliskeVariabilityLocationHigh2018}
Stephen~V. Gliske, Zachary~T. Irwin, Cynthia Chestek, Garnett~L. Hegeman,
  Benjamin Brinkmann, Oren Sagher, Hugh J.~L. Garton, Greg~A. Worrell, and
  William~C. Stacey.
\newblock Variability in the location of high frequency oscillations during
  prolonged intracranial {{EEG}} recordings.
\newblock \emph{Nature Communications}, 9\penalty0 (1):\penalty0 2155, December
  2018.
\newblock ISSN 2041-1723.
\newblock \doi{10.1038/s41467-018-04549-2}.
\newblock URL \url{http://www.nature.com/articles/s41467-018-04549-2}.

\bibitem[Huang et~al.(1998)Huang, Shen, Long, Wu, Shih, Zheng, Yen, Tung, and
  Liu]{huangEmpiricalModeDecomposition1998}
Norden~E. Huang, Zheng Shen, Steven~R. Long, Manli~C. Wu, Hsing~H. Shih, Quanan
  Zheng, Nai-Chyuan Yen, Chi~Chao Tung, and Henry~H. Liu.
\newblock The empirical mode decomposition and the {{Hilbert}} spectrum for
  nonlinear and non-stationary time series analysis.
\newblock \emph{Proceedings of the Royal Society of London. Series A:
  Mathematical, Physical and Engineering Sciences}, 454\penalty0
  (1971):\penalty0 903--995, March 1998.
\newblock ISSN 1364-5021, 1471-2946.
\newblock \doi{10.1098/rspa.1998.0193}.
\newblock URL
  \url{https://royalsocietypublishing.org/doi/10.1098/rspa.1998.0193}.

\bibitem[Huang et~al.(2003)Huang, Wu, Long, Shen, Qu, Gloersen, and
  Fan]{huangConfidenceLimitEmpirical2003}
Norden~E Huang, Man-Li~C Wu, Steven~R Long, Samuel~S.P Shen, Wendong Qu, Per
  Gloersen, and Kuang~L Fan.
\newblock A confidence limit for the empirical mode decomposition and
  {{Hilbert}} spectral analysis.
\newblock \emph{Proceedings of the Royal Society of London. Series A:
  Mathematical, Physical and Engineering Sciences}, 459\penalty0
  (2037):\penalty0 2317--2345, September 2003.
\newblock ISSN 1364-5021, 1471-2946.
\newblock \doi{10.1098/rspa.2003.1123}.
\newblock URL
  \url{https://royalsocietypublishing.org/doi/10.1098/rspa.2003.1123}.

\bibitem[Huang(2014)]{huangHilbertHuangTransformIts2014}
Norden~Eh Huang.
\newblock \emph{Hilbert-{{Huang Transform}} and {{Its Applications}}}.
\newblock {World Scientific}, 2014.
\newblock ISBN 978-981-4508-24-7.

\bibitem[Ilyas et~al.(2022)Ilyas, Toth, Chaitanya, Riley, and
  Pati]{ilyasIctalHighfrequencyActivity2022}
Adeel Ilyas, Emilia Toth, Ganne Chaitanya, Kristen Riley, and Sandipan Pati.
\newblock Ictal high-frequency activity in limbic thalamic nuclei varies with
  electrographic seizure-onset patterns in temporal lobe epilepsy.
\newblock \emph{Clinical Neurophysiology}, 137:\penalty0 183--192, May 2022.
\newblock ISSN 1388-2457.
\newblock \doi{10.1016/j.clinph.2022.01.134}.
\newblock URL
  \url{https://www.sciencedirect.com/science/article/pii/S1388245722001596}.

\bibitem[Jenssen et~al.(2006)Jenssen, Gracely, and
  Sperling]{jenssenHowLongMost2006}
Sigmund Jenssen, Edward~J. Gracely, and Michael~R. Sperling.
\newblock How {{Long Do Most Seizures Last}}? {{A Systematic Comparison}} of
  {{Seizures Recorded}} in the {{Epilepsy Monitoring Unit}}.
\newblock \emph{Epilepsia}, 47\penalty0 (9):\penalty0 1499--1503, 2006.
\newblock ISSN 1528-1167.
\newblock \doi{10.1111/j.1528-1167.2006.00622.x}.
\newblock URL
  \url{https://onlinelibrary.wiley.com/doi/abs/10.1111/j.1528-1167.2006.00622.x}.

\bibitem[Jin et~al.(2020)Jin, Aung, Geng, and
  Wang]{jinEpilepsyItsInteraction2020a}
Bo~Jin, Thandar Aung, Yu~Geng, and Shuang Wang.
\newblock Epilepsy and {{Its Interaction With Sleep}} and {{Circadian Rhythm}}.
\newblock \emph{Frontiers in Neurology}, 11:\penalty0 327, May 2020.
\newblock ISSN 1664-2295.
\newblock \doi{10.3389/fneur.2020.00327}.
\newblock URL
  \url{https://www.frontiersin.org/article/10.3389/fneur.2020.00327/full}.

\bibitem[Jobst et~al.(2001)Jobst, Williamson, Neuschwander, Darcey, Thadani,
  and Roberts]{jobstSecondarilyGeneralizedSeizures2001}
Barbara~C. Jobst, Peter~D. Williamson, Timothy~B. Neuschwander, Terrance~M.
  Darcey, Vijay~M. Thadani, and David~W. Roberts.
\newblock Secondarily {{Generalized Seizures}} in {{Mesial Temporal Epilepsy}}:
  {{Clinical Characteristics}}, {{Lateralizing Signs}}, and {{Association With
  Sleep}}\textendash{{Wake Cycle}}.
\newblock \emph{Epilepsia}, 42\penalty0 (10):\penalty0 1279--1287, October
  2001.
\newblock ISSN 0013-9580, 1528-1167.
\newblock \doi{10.1046/j.1528-1157.2001.09701.x}.
\newblock URL
  \url{https://onlinelibrary.wiley.com/doi/abs/10.1046/j.1528-1157.2001.09701.x}.

\bibitem[Kaplan et~al.(2005)Kaplan, Fingelkurts, Fingelkurts, Borisov, and
  Darkhovsky]{kaplanNonstationaryNatureBrain2005}
Alexander~Ya. Kaplan, Andrew~A. Fingelkurts, Alexander~A. Fingelkurts,
  Sergei~V. Borisov, and Boris~S. Darkhovsky.
\newblock Nonstationary nature of the brain activity as revealed by
  {{EEG}}/{{MEG}}: {{Methodological}}, practical and conceptual challenges.
\newblock \emph{Signal Processing}, 85\penalty0 (11):\penalty0 2190--2212,
  November 2005.
\newblock ISSN 01651684.
\newblock \doi{10.1016/j.sigpro.2005.07.010}.
\newblock URL
  \url{https://linkinghub.elsevier.com/retrieve/pii/S0165168405002094}.

\bibitem[Karoly et~al.(2016)Karoly, Freestone, Boston, Grayden, Himes, Leyde,
  Seneviratne, Berkovic, O'Brien, and
  Cook]{karolyInterictalSpikesEpileptic2016}
Philippa~J. Karoly, Dean~R. Freestone, Ray Boston, David~B. Grayden, David
  Himes, Kent Leyde, Udaya Seneviratne, Samuel Berkovic, Terence O'Brien, and
  Mark~J. Cook.
\newblock Interictal spikes and epileptic seizures: Their relationship and
  underlying rhythmicity.
\newblock \emph{Brain}, 139\penalty0 (4):\penalty0 1066--1078, April 2016.
\newblock ISSN 0006-8950.
\newblock \doi{10.1093/brain/aww019}.
\newblock URL \url{http://academic.oup.com/brain/article/139/4/1066/2464379}.

\bibitem[Karoly et~al.(2017)Karoly, Ung, Grayden, Kuhlmann, Leyde, Cook, and
  Freestone]{karolyCircadianProfileEpilepsy2017}
Philippa~J. Karoly, Hoameng Ung, David~B. Grayden, Levin Kuhlmann, Kent Leyde,
  Mark~J. Cook, and Dean~R. Freestone.
\newblock The circadian profile of epilepsy improves seizure forecasting.
\newblock \emph{Brain}, 140\penalty0 (8):\penalty0 2169--2182, August 2017.
\newblock ISSN 0006-8950.
\newblock \doi{10.1093/brain/awx173}.
\newblock URL \url{https://academic.oup.com/brain/article/140/8/2169/4032453}.

\bibitem[Karoly et~al.(2018)Karoly, Goldenholz, Freestone, Moss, Grayden,
  Theodore, and Cook]{karolyCircadianCircaseptanRhythms2018}
Philippa~J Karoly, Daniel~M Goldenholz, Dean~R Freestone, Robert~E Moss,
  David~B Grayden, William~H Theodore, and Mark~J Cook.
\newblock Circadian and circaseptan rhythms in human epilepsy: A retrospective
  cohort study.
\newblock \emph{The Lancet Neurology}, 17\penalty0 (11):\penalty0 977--985,
  November 2018.
\newblock ISSN 1474-4422.
\newblock \doi{10.1016/S1474-4422(18)30274-6}.
\newblock URL
  \url{http://www.sciencedirect.com/science/article/pii/S1474442218302746}.

\bibitem[Karoly et~al.(2021)Karoly, Rao, Gregg, Worrell, Bernard, Cook, and
  Baud]{karolyCyclesEpilepsy2021}
Philippa~J. Karoly, Vikram~R. Rao, Nicholas~M. Gregg, Gregory~A. Worrell,
  Christophe Bernard, Mark~J. Cook, and Maxime~O. Baud.
\newblock Cycles in epilepsy.
\newblock \emph{Nature Reviews Neurology}, 17\penalty0 (5):\penalty0 267--284,
  May 2021.
\newblock ISSN 1759-4766.
\newblock \doi{10.1038/s41582-021-00464-1}.
\newblock URL \url{http://www.nature.com/articles/s41582-021-00464-1}.

\bibitem[Karthick et~al.(2018)Karthick, Tanaka, Khoo, and
  Gotman]{karthickPredictionSecondaryGeneralization2018}
P.~A. Karthick, Hideaki Tanaka, Hui~Ming Khoo, and Jean Gotman.
\newblock Prediction of secondary generalization from a focal onset seizure in
  intracerebral {{EEG}}.
\newblock \emph{Clinical Neurophysiology}, 129\penalty0 (5):\penalty0
  1030--1040, May 2018.
\newblock ISSN 1388-2457.
\newblock \doi{10.1016/j.clinph.2018.02.122}.
\newblock URL
  \url{https://www.sciencedirect.com/science/article/pii/S1388245718302311}.

\bibitem[Kaufmann et~al.(2020)Kaufmann, Seethaler, Lauseker, Fan, Vollmar,
  Noachtar, and R{\'e}mi]{kaufmannWhoSeizesLongest2020}
Elisabeth Kaufmann, Magdalena Seethaler, Michael Lauseker, Min Fan, Christian
  Vollmar, Soheyl Noachtar, and Jan R{\'e}mi.
\newblock Who seizes longest? {{Impact}} of clinical and demographic factors.
\newblock \emph{Epilepsia}, 61\penalty0 (7):\penalty0 1376--1385, July 2020.
\newblock ISSN 0013-9580, 1528-1167.
\newblock \doi{10.1111/epi.16577}.
\newblock URL \url{https://onlinelibrary.wiley.com/doi/10.1111/epi.16577}.

\bibitem[Kim et~al.(2011)Kim, Cho, Lee, Joo, Hong, Hong, and
  Seo]{kimSeizureDurationDetermined2011}
Daeyoung Kim, Jae-Wook Cho, Jihyun Lee, Eun~Yeon Joo, Seung~Chyul Hong,
  Seung~Bong Hong, and Dae-Won Seo.
\newblock Seizure {{Duration Determined}} by {{Subdural Electrode Recordings}}
  in {{Adult Patients}} with {{Intractable Focal Epilepsy}}.
\newblock \emph{Journal of Epilepsy Research}, 1\penalty0 (2):\penalty0 57--64,
  December 2011.
\newblock ISSN 2233-6249.
\newblock \doi{10.14581/jer.11011}.
\newblock URL \url{https://www.ncbi.nlm.nih.gov/pmc/articles/PMC3952333/}.

\bibitem[Kirby et~al.(2020)Kirby, Leach, Brockington, Patsalos, Reuber, and
  Leach]{kirbyDrugWithdrawalEpilepsy2020}
Jack Kirby, Veronica~M Leach, Alice Brockington, Phillip Patsalos, Markus
  Reuber, and John~Paul Leach.
\newblock Drug withdrawal in the epilepsy monitoring unit \textendash{} {{The}}
  patsalos table.
\newblock \emph{Seizure}, 75:\penalty0 75--81, February 2020.
\newblock ISSN 1059-1311.
\newblock \doi{10.1016/j.seizure.2019.12.010}.
\newblock URL
  \url{https://www.sciencedirect.com/science/article/pii/S1059131119307344}.

\bibitem[Lamberts et~al.(2013)Lamberts, Gaitatzis, Sander, Elger, Surges, and
  Thijs]{lambertsPostictalGeneralizedEEG2013}
Robert~J. Lamberts, Athanasios Gaitatzis, Josemir~W. Sander, Christian~E.
  Elger, Rainer Surges, and Roland~D. Thijs.
\newblock Postictal generalized {{EEG}} suppression.
\newblock \emph{Neurology}, 81\penalty0 (14):\penalty0 1252--1256, October
  2013.
\newblock ISSN 0028-3878.
\newblock \doi{10.1212/WNL.0b013e3182a6cbeb}.
\newblock URL \url{https://www.ncbi.nlm.nih.gov/pmc/articles/PMC3795608/}.

\bibitem[Lehnertz et~al.(2017)Lehnertz, Geier, Rings, and
  Stahn]{lehnertzCapturingTimevaryingBrain2017}
Klaus Lehnertz, Christian Geier, Thorsten Rings, and Kirsten Stahn.
\newblock Capturing time-varying brain dynamics.
\newblock \emph{EPJ Nonlinear Biomedical Physics}, 5:\penalty0 2, 2017.
\newblock ISSN 2195-0008.
\newblock \doi{10.1051/epjnbp/2017001}.
\newblock URL \url{http://www.epj-nbp.org/10.1051/epjnbp/2017001}.

\bibitem[Loddenkemper et~al.(2011)Loddenkemper, Vendrame, Zarowski, Gregas,
  Alexopoulos, Wyllie, and Kothare]{loddenkemperCircadianPatternsPediatric2011}
T.~Loddenkemper, M.~Vendrame, M.~Zarowski, M.~Gregas, A.~V. Alexopoulos,
  E.~Wyllie, and S.~V. Kothare.
\newblock Circadian patterns of pediatric seizures.
\newblock \emph{Neurology}, 76\penalty0 (2):\penalty0 145--153, January 2011.
\newblock ISSN 1526-632X.
\newblock \doi{10.1212/WNL.0b013e318206ca46}.

\bibitem[Mardia(1976)]{mardiaLinearCircularCorrelationCoefficients1976}
K.~V. Mardia.
\newblock Linear-{{Circular Correlation Coefficients}} and {{Rhythmometry}}.
\newblock \emph{Biometrika}, 63\penalty0 (2):\penalty0 403--405, 1976.
\newblock ISSN 0006-3444.
\newblock \doi{10.2307/2335637}.
\newblock URL \url{http://www.jstor.org/stable/2335637}.

\bibitem[Mardia and Jupp(1999)]{mardiaDirectionalStatistics1999}
Kanti~V. Mardia and Peter~E. Jupp, editors.
\newblock \emph{Directional {{Statistics}}}.
\newblock Wiley {{Series}} in {{Probability}} and {{Statistics}}. {John Wiley
  \& Sons, Inc.}, {Hoboken, NJ, USA}, January 1999.
\newblock ISBN 978-0-470-31697-9 978-0-471-95333-3.
\newblock \doi{10.1002/9780470316979}.
\newblock URL \url{http://doi.wiley.com/10.1002/9780470316979}.

\bibitem[Mitsis et~al.(2020)Mitsis, Anastasiadou, Christodoulakis,
  Papathanasiou, Papacostas, and Hadjipapas]{mitsisFunctionalBrainNetworks2020}
Georgios~D. Mitsis, Maria~N. Anastasiadou, Manolis Christodoulakis,
  Eleftherios~S. Papathanasiou, Savvas~S. Papacostas, and Avgis Hadjipapas.
\newblock Functional brain networks of patients with epilepsy exhibit
  pronounced multiscale periodicities, which correlate with seizure onset.
\newblock \emph{Human Brain Mapping}, 41\penalty0 (8):\penalty0 2059--2076,
  2020.
\newblock ISSN 1097-0193.
\newblock \doi{10.1002/hbm.24930}.
\newblock URL \url{https://onlinelibrary.wiley.com/doi/abs/10.1002/hbm.24930}.

\bibitem[{Ochoa-Urrea} et~al.(2021){Ochoa-Urrea}, Lacuey, Vilella, Zhu,
  {Jamal-Omidi}, Rani, Hampson, Dayyani, Hampson, Hupp, Tao, Sainju, Friedman,
  Nei, Scott, Allen, Gehlbach, {Reick-Mitrisin}, Schuele, Ogren, Harper, Diehl,
  Bateman, Devinsky, Richerson, Zhang, and
  Lhatoo]{ochoa-urreaSeizureClustersSeizure2021}
Manuela {Ochoa-Urrea}, Nuria Lacuey, Laura Vilella, Liang Zhu, Shirin
  {Jamal-Omidi}, M.~R.~Sandhya Rani, Johnson~P. Hampson, Mojtaba Dayyani,
  Jaison Hampson, Norma~J. Hupp, Shiqiang Tao, Rup~K. Sainju, Daniel Friedman,
  Maromi Nei, Catherine Scott, Luke Allen, Brian~K. Gehlbach, Victoria
  {Reick-Mitrisin}, Stephan Schuele, Jennifer Ogren, Ronald~M. Harper, Beate
  Diehl, Lisa~M. Bateman, Orrin Devinsky, George~B. Richerson, Guo-Qiang Zhang,
  and Samden~D. Lhatoo.
\newblock Seizure {{Clusters}}, {{Seizure Severity Markers}}, and {{SUDEP
  Risk}}.
\newblock \emph{Frontiers in Neurology}, 12, 2021.
\newblock ISSN 1664-2295.
\newblock URL
  \url{https://www.frontiersin.org/articles/10.3389/fneur.2021.643916}.

\bibitem[Panagiotopoulou et~al.(2022)Panagiotopoulou, Papasavvas, Schroeder,
  Thomas, Taylor, and Wang]{panagiotopoulouFluctuationsEEGBand2022}
M.~Panagiotopoulou, C.~A. Papasavvas, G.~M. Schroeder, R.~H. Thomas, P.~N.
  Taylor, and Y.~Wang.
\newblock Fluctuations in {{EEG}} band power at subject-specific timescales
  over minutes to days explain changes in seizure evolutions.
\newblock \emph{Human Brain Mapping}, 2022.
\newblock \doi{10.1002/hbm.25796}.
\newblock URL \url{https://eprints.ncl.ac.uk}.

\bibitem[Pandolfo(2015)]{pandolfoCircularStatistics2015}
Giuseppe Pandolfo.
\newblock Circular statistics in {{R}}.
\newblock \emph{Journal of Applied Statistics}, 42\penalty0 (4):\penalty0
  918--919, April 2015.
\newblock ISSN 0266-4763, 1360-0532.
\newblock \doi{10.1080/02664763.2014.991072}.
\newblock URL
  \url{http://www.tandfonline.com/doi/abs/10.1080/02664763.2014.991072}.

\bibitem[Pattnaik et~al.(2022)Pattnaik, Ghosn, Ong, Revell, Ojemann, Scheid,
  Bernabei, Conrad, Sinha, Davis, Sinha, and
  Litt]{pattnaikQuantitativeToolSeizure2022a}
Akash~R. Pattnaik, Nina~J. Ghosn, Ian~Z. Ong, Andrew~Y. Revell, William K.~S.
  Ojemann, Brittany~H. Scheid, John~M. Bernabei, Erin Conrad, Saurabh~R. Sinha,
  Kathryn~A. Davis, Nishant Sinha, and Brian Litt.
\newblock A quantitative tool for seizure severity: Diagnostic and therapeutic
  applications, November 2022.
\newblock URL
  \url{https://www.medrxiv.org/content/10.1101/2022.10.26.22281569v1}.

\bibitem[Payne et~al.(2018)Payne, Karoly, Freestone, Boston, D'Souza, Nurse,
  Kuhlmann, Cook, and Grayden]{paynePostictalSuppressionSeizure2018}
Daniel~E. Payne, Philippa~J. Karoly, Dean~R. Freestone, Ray Boston, Wendyl
  D'Souza, Ewan Nurse, Levin Kuhlmann, Mark~J. Cook, and David~B. Grayden.
\newblock Postictal suppression and seizure durations: {{A}} patient-specific,
  long-term {{iEEG}} analysis.
\newblock \emph{Epilepsia}, 59\penalty0 (5):\penalty0 1027--1036, 2018.
\newblock ISSN 1528-1167.
\newblock \doi{10.1111/epi.14065}.
\newblock URL \url{http://onlinelibrary.wiley.com/doi/abs/10.1111/epi.14065}.

\bibitem[Payne et~al.(2021)Payne, Dell, Karoly, Kremen, Gerla, Kuhlmann,
  Worrell, Cook, Grayden, and Freestone]{payneIdentifyingSeizureRisk2021}
Daniel~E. Payne, Katrina~L. Dell, Phillipa~J. Karoly, Vaclav Kremen, Vaclav
  Gerla, Levin Kuhlmann, Gregory~A. Worrell, Mark~J. Cook, David~B. Grayden,
  and Dean~R. Freestone.
\newblock Identifying seizure risk factors: {{A}} comparison of sleep, weather,
  and temporal features using a {{Bayesian}} forecast.
\newblock \emph{Epilepsia}, 62\penalty0 (2):\penalty0 371--382, February 2021.
\newblock ISSN 0013-9580, 1528-1167.
\newblock \doi{10.1111/epi.16785}.
\newblock URL \url{https://onlinelibrary.wiley.com/doi/10.1111/epi.16785}.

\bibitem[Peng et~al.(2017)Peng, Danison, and
  Seyal]{pengPostictalGeneralizedEEG2017}
Weifeng Peng, Jessica~L. Danison, and Masud Seyal.
\newblock Postictal generalized {{EEG}} suppression and respiratory dysfunction
  following generalized tonic\textendash clonic seizures in sleep and
  wakefulness.
\newblock \emph{Epilepsia}, 58\penalty0 (8):\penalty0 1409--1414, 2017.
\newblock ISSN 1528-1167.
\newblock \doi{10.1111/epi.13805}.
\newblock URL \url{http://onlinelibrary.wiley.com/doi/abs/10.1111/epi.13805}.

\bibitem[Pewsey and
  {Garc{\'i}a-Portugu{\'e}s}(2020)]{pewseyRecentAdvancesDirectional2020}
Arthur Pewsey and Eduardo {Garc{\'i}a-Portugu{\'e}s}.
\newblock Recent advances in directional statistics.
\newblock \emph{arXiv:2005.06889 [stat]}, September 2020.
\newblock URL \url{http://arxiv.org/abs/2005.06889}.

\bibitem[Proix et~al.(2021)Proix, Truccolo, Leguia, Tcheng, {King-Stephens},
  Rao, and Baud]{proixForecastingSeizureRisk2021}
Timoth{\'e}e Proix, Wilson Truccolo, Marc~G. Leguia, Thomas~K. Tcheng, David
  {King-Stephens}, Vikram~R. Rao, and Maxime~O. Baud.
\newblock Forecasting seizure risk in adults with focal epilepsy: A development
  and validation study.
\newblock \emph{The Lancet. Neurology}, 20\penalty0 (2):\penalty0 127--135,
  February 2021.
\newblock ISSN 1474-4465.
\newblock \doi{10.1016/S1474-4422(20)30396-3}.

\bibitem[Ramgopal et~al.(2013)Ramgopal, {Thome-Souza}, and
  Loddenkemper]{ramgopalChronopharmacologyAntiConvulsiveTherapy2013}
Sriram Ramgopal, Sigride {Thome-Souza}, and Tobias Loddenkemper.
\newblock Chronopharmacology of {{Anti-Convulsive Therapy}}.
\newblock \emph{Current Neurology and Neuroscience Reports}, 13\penalty0
  (4):\penalty0 339, March 2013.
\newblock ISSN 1534-6293.
\newblock \doi{10.1007/s11910-013-0339-2}.
\newblock URL \url{https://doi.org/10.1007/s11910-013-0339-2}.

\bibitem[Rehman and Mandic(2010)]{rehmanMultivariateEmpiricalMode2010}
N.~Rehman and D.~P. Mandic.
\newblock Multivariate empirical mode decomposition.
\newblock \emph{Proceedings of the Royal Society A: Mathematical, Physical and
  Engineering Sciences}, 466\penalty0 (2117):\penalty0 1291--1302, May 2010.
\newblock \doi{10.1098/rspa.2009.0502}.
\newblock URL
  \url{https://royalsocietypublishing.org/doi/10.1098/rspa.2009.0502}.

\bibitem[Rilling et~al.(2007)Rilling, Flandrin, Goncalves, and
  Lilly]{rillingBivariateEmpiricalMode2007}
Gabriel Rilling, Patrick Flandrin, Paulo Goncalves, and Jonathan~M. Lilly.
\newblock Bivariate {{Empirical Mode Decomposition}}.
\newblock \emph{IEEE Signal Processing Letters}, 14\penalty0 (12):\penalty0
  936--939, December 2007.
\newblock ISSN 1070-9908, 1558-2361.
\newblock \doi{10.1109/LSP.2007.904710}.

\bibitem[Schroeder et~al.(2020)Schroeder, Diehl, Chowdhury, Duncan, de~Tisi,
  Trevelyan, Forsyth, Jackson, Taylor, and
  Wang]{schroederSeizurePathwaysChange2020}
Gabrielle~M. Schroeder, Beate Diehl, Fahmida~A. Chowdhury, John~S. Duncan, Jane
  de~Tisi, Andrew~J. Trevelyan, Rob Forsyth, Andrew Jackson, Peter~N. Taylor,
  and Yujiang Wang.
\newblock Seizure pathways change on circadian and slower timescales in
  individual patients with focal epilepsy.
\newblock \emph{Proceedings of the National Academy of Sciences}, May 2020.
\newblock ISSN 0027-8424, 1091-6490.
\newblock \doi{10.1073/pnas.1922084117}.
\newblock URL \url{https://www.pnas.org/content/early/2020/05/01/1922084117}.

\bibitem[Schroeder et~al.(2022{\natexlab{a}})Schroeder, Chowdhury, Cook, Diehl,
  Duncan, Karoly, Taylor, and Wang]{schroederMultipleMechanismsShape2022}
Gabrielle~M Schroeder, Fahmida~A Chowdhury, Mark~J Cook, Beate Diehl, John~S
  Duncan, Philippa~J Karoly, Peter~N Taylor, and Yujiang Wang.
\newblock Multiple mechanisms shape the relationship between pathway and
  duration of focal seizures.
\newblock \emph{Brain Communications}, 4\penalty0 (4):\penalty0 fcac173, August
  2022{\natexlab{a}}.
\newblock ISSN 2632-1297.
\newblock \doi{10.1093/braincomms/fcac173}.
\newblock URL \url{https://doi.org/10.1093/braincomms/fcac173}.

\bibitem[Schroeder et~al.(2022{\natexlab{b}})Schroeder, Karoly, Maturana,
  Taylor, Cook, and Wang]{schroederChronicIEEGRecordings2022}
Gabrielle~M. Schroeder, Philippa~J. Karoly, Matias Maturana, Peter~N. Taylor,
  Mark~J. Cook, and Yujiang Wang.
\newblock Chronic {{iEEG}} recordings and interictal spike rate reveal
  multiscale temporal modulations in seizure states, January
  2022{\natexlab{b}}.
\newblock URL \url{http://arxiv.org/abs/2201.11600}.

\bibitem[Silva et~al.(2003)Silva, Blanes, Kalitzin, Parra, Suffczynski, and
  Velis]{silvaEpilepsiesDynamicalDiseases2003}
Fernando Lopes~Da Silva, Wouter Blanes, Stiliyan~N. Kalitzin, Jaime Parra,
  Piotr Suffczynski, and Demetrios~N. Velis.
\newblock Epilepsies as {{Dynamical Diseases}} of {{Brain Systems}}: {{Basic
  Models}} of the {{Transition Between Normal}} and {{Epileptic Activity}}.
\newblock \emph{Epilepsia}, 44\penalty0 (s12):\penalty0 72--83, 2003.
\newblock ISSN 1528-1167.
\newblock \doi{10.1111/j.0013-9580.2003.12005.x}.
\newblock URL
  \url{https://onlinelibrary.wiley.com/doi/abs/10.1111/j.0013-9580.2003.12005.x}.

\bibitem[Simeone et~al.(2018)Simeone, Simeone, Stafstrom, and
  Rho]{simeoneKetoneBodiesMediate2018}
Timothy~A. Simeone, Kristina~A. Simeone, Carl~E. Stafstrom, and Jong~M. Rho.
\newblock Do ketone bodies mediate the anti-seizure effects of the ketogenic
  diet?
\newblock \emph{Neuropharmacology}, 133:\penalty0 233--241, May 2018.
\newblock ISSN 0028-3908.
\newblock \doi{10.1016/j.neuropharm.2018.01.011}.
\newblock URL
  \url{https://www.sciencedirect.com/science/article/pii/S002839081830011X}.

\bibitem[Sinha et~al.(2006)Sinha, Brady, Scott, and
  Walker]{sinhaSeizuresPatientsRefractory2006}
S~Sinha, M~Brady, C~A Scott, and M~C Walker.
\newblock Do seizures in patients with refractory epilepsy vary between
  wakefulness and sleep?
\newblock \emph{Journal of Neurology, Neurosurgery, and Psychiatry},
  77\penalty0 (9):\penalty0 1076--1078, September 2006.
\newblock ISSN 0022-3050.
\newblock \doi{10.1136/jnnp.2006.088385}.
\newblock URL \url{https://www.ncbi.nlm.nih.gov/pmc/articles/PMC2077754/}.

\bibitem[Spencer et~al.(2016)Spencer, Sun, Brown, Jobst, Fountain, Wong, Mirro,
  and Quigg]{spencerCircadianUltradianPatterns2016}
David~C. Spencer, Felice~T. Sun, Sarah~N. Brown, Barbara~C. Jobst, Nathan~B.
  Fountain, Victoria S.~S. Wong, Emily~A. Mirro, and Mark Quigg.
\newblock Circadian and ultradian patterns of epileptiform discharges differ by
  seizure-onset location during long-term ambulatory intracranial monitoring.
\newblock \emph{Epilepsia}, 57\penalty0 (9):\penalty0 1495--1502, 2016.
\newblock ISSN 1528-1167.
\newblock \doi{10.1111/epi.13455}.
\newblock URL \url{https://onlinelibrary.wiley.com/doi/abs/10.1111/epi.13455}.

\bibitem[Spencer et~al.(1981)Spencer, Spencer, Williamson, and
  Mattson]{spencerIctalEffectsAnticonvulsant1981}
Susan~S. Spencer, Dennis~D. Spencer, P{\'e}t{\'e}r~D. Williamson, and
  Richard~H. Mattson.
\newblock Ictal {{Effects}} of {{Anticonvulsant Medication Withdrawal}} in
  {{Epileptic Patients}}.
\newblock \emph{Epilepsia}, 22\penalty0 (3):\penalty0 297--307, 1981.
\newblock ISSN 1528-1167.
\newblock \doi{10.1111/j.1528-1157.1981.tb04113.x}.
\newblock URL
  \url{http://onlinelibrary.wiley.com/doi/abs/10.1111/j.1528-1157.1981.tb04113.x}.

\bibitem[Suffczynski et~al.(2006)Suffczynski, {Lopes da Silva}, Parra, Velis,
  Bouwman, {van Rijn}, {van Hese}, Boon, Khosravani, Derchansky, Carlen, and
  Kalitzin]{suffczynskiDynamicsEpilepticPhenomena2006}
P.~Suffczynski, F.H. {Lopes da Silva}, J.~Parra, D.N. Velis, B.M. Bouwman, C.M.
  {van Rijn}, P.~{van Hese}, P.~Boon, H.~Khosravani, M.~Derchansky, P.~Carlen,
  and S.~Kalitzin.
\newblock Dynamics of {{Epileptic Phenomena Determined From Statistics}} of
  {{Ictal Transitions}}.
\newblock \emph{IEEE Transactions on Biomedical Engineering}, 53\penalty0
  (3):\penalty0 524--532, March 2006.
\newblock ISSN 0018-9294.
\newblock \doi{10.1109/TBME.2005.869800}.
\newblock URL \url{http://ieeexplore.ieee.org/document/1597503/}.

\bibitem[Todorova et~al.(2013)Todorova, Velikova, Kaprelyan, and
  Tsekov]{todorovaSEIZURESEVERITYALTERNATIVE2013}
Koraliya~S. Todorova, Valentina~S. Velikova, Ara~G. Kaprelyan, and Stefan~T.
  Tsekov.
\newblock {{SEIZURE SEVERITY AS AN ALTERNATIVE MEASURE OF OUTCOME IN
  EPILEPSY}}.
\newblock \emph{Journal of IMAB - Annual Proceeding (Scientific Papers)},
  19\penalty0 (3):\penalty0 433--437, July 2013.
\newblock ISSN 1312773X.
\newblock \doi{10.5272/jimab.2013193.433}.
\newblock URL
  \url{http://www.journal-imab-bg.org/issue-2013/issue3/vol19book3p433-437.html}.

\bibitem[Verbeek et~al.(2016)Verbeek, Leen, Willemsen, Slats, and
  Claassen]{verbeekHourlyAnalysisCerebrospinal2016}
Marcel~M Verbeek, Wilhelmina~G Leen, Mich{\`e}l~A Willemsen, Diane Slats, and
  Jurgen~A Claassen.
\newblock Hourly analysis of cerebrospinal fluid glucose shows large diurnal
  fluctuations.
\newblock \emph{Journal of Cerebral Blood Flow \& Metabolism}, 36\penalty0
  (5):\penalty0 899--902, May 2016.
\newblock ISSN 0271-678X.
\newblock \doi{10.1177/0271678X16637612}.
\newblock URL \url{https://www.ncbi.nlm.nih.gov/pmc/articles/PMC4853846/}.

\bibitem[Yuan and Lin(2006)]{yuanModelSelectionEstimation2006}
Ming Yuan and Yi~Lin.
\newblock Model selection and estimation in regression with grouped variables.
\newblock \emph{Journal of the Royal Statistical Society: Series B (Statistical
  Methodology)}, 68\penalty0 (1):\penalty0 49--67, February 2006.
\newblock ISSN 1369-7412, 1467-9868.
\newblock \doi{10.1111/j.1467-9868.2005.00532.x}.
\newblock URL
  \url{https://onlinelibrary.wiley.com/doi/10.1111/j.1467-9868.2005.00532.x}.

\bibitem[Zhou et~al.(2002)Zhou, Wang, Hopp, Kerling, Kirchner, Pauli, and
  Stefan]{zhouInfluenceIctalSeizure2002}
Dong Zhou, Ying Wang, Peter Hopp, Frank Kerling, Annette Kirchner, Elisabeth
  Pauli, and Hermann Stefan.
\newblock Influence on {{Ictal Seizure Semiology}} of {{Rapid Withdrawal}} of
  {{Carbamazepine}} and {{Valproate}} in {{Monotherapy}}.
\newblock \emph{Epilepsia}, 43\penalty0 (4):\penalty0 386--393, 2002.
\newblock ISSN 1528-1167.
\newblock \doi{10.1046/j.1528-1157.2002.45201.x}.
\newblock URL
  \url{https://onlinelibrary.wiley.com/doi/abs/10.1046/j.1528-1157.2002.45201.x}.

\end{thebibliography}
\newpage

%%%%%%%%%% %%%%%%%%%% SUPPLEMENTARY %%%%%%%%%% %%%%%%%%%% 

\renewcommand{\thefigure}{S\arabic{figure}}
\setcounter{figure}{0}
\counterwithin{figure}{section}
\counterwithin{table}{section}
\renewcommand\thesection{S\arabic{section}}
\setcounter{section}{0}
%%%%%%%%%%%%%%%%%%%%%%%%%%%%%%%%%%%%%%%%%%%%%%%%%%%
%%%%%%%%%%%%%%%%%%%%%%%%%%%%%%%%%%%%%%%%%%%%%%%%%%%
%%%%%%%       Supplementary
%%%%%%%%%%%%%%%%%%%%%%%%%%%%%%%%%%%%%%%%%%%%%%%%%%%%
%%%%%%%%%%%%%%%%%%%%%%%%%%%%%%%%%%%%%%%%%%%%%%%%%%%%

\section*{Supplementary}

\subsection{Imputation of missing data\label{suppl:ImputeMissing}}

To allow subsequent steps of analysis, we imputed any missing data in the band power matrix $A$ before extracting the cycles on different timescales in the data. Missing values were later placed at the corresponding entries in the data for the final output (i.e. we are not using imputed data for our final analysis). For each frequency band, we identified all the missing blocks of the band power and further imputed them. For missing blocks of size equal to one, we used the mean of one value before and after the missing block. For missing blocks of size greater than one, we first identified the segments of equal length with the missing block, before and after the missing block. In cases where the missing blocks were at the start of the recording or the preceding segment was lower in size than the missing block, we applied imputation using just the segment following the missing block. We linearly interpolated the data of missing blocks using the mean of the surrounding segments. The final imputed values were the interpolated ones with Gaussian noise of mean zero and standard deviation, the 60\% of the standard deviation of the surrounding segments. If there were missing data apparent in the surrounding segments, those were ignored. 

\subsection{Capturing oscillatory modes using MEMD\label{suppl:MEMD_theory}}

In order to extract oscillatory modes embedded in band power, we applied a signal decomposition method called Empirical Mode Decomposition (EMD)~\citep{huangEmpiricalModeDecomposition1998, huangConfidenceLimitEmpirical2003}. EMD does not require any assumption about stationary or linear characteristics of the signal; EEG signals are non-stationary processes influenced by complex non-linear dynamics~\citep{kaplanNonstationaryNatureBrain2005, fingelkurtsOperationalArchitectonicsHuman2001, lehnertzCapturingTimevaryingBrain2017}. 
EMD captures a finite number, $M$ of narrow-band modes from an input signal $Y(t)$, known as intrinsic mode functions (IMFs), based on the local minima and maxima of the signal: $Y(t)=\sum_{i=1}^M \text{IMF}_i(t) + r(t)$, where $r(t)$ is the residue signal~\citep{huangEmpiricalModeDecomposition1998}. 
The IMFs correspond to a limited-band frequency and have a local mean of zero, satisfying the properties needed for Hilbert-transform to be well-defined. Thus, Hilbert analysis can be applied to each IMF yielding instantaneous characteristics of the signals, such as amplitude, frequency and phases.

Therefore, in order to obtain a time-frequency representation of the oscillatory modes (IMFs), and hence derive their time-varying characteristics (instantaneous frequency, phase, and amplitude), we applied a Hilbert-transform on each dimension of the IMF (following Hilbert Spectral Analysis methods for EMD)~\citep{huangEmpiricalModeDecomposition1998, huangConfidenceLimitEmpirical2003, huangHilbertHuangTransformIts2014}. %We will refer to these oscillatory modes as "band power IMFs" throughout the text. 

However, in order to decompose multivariate signals (in our case band power across frequency bands) we used an extension of the EMD to multi-dimensional space, called the Multivariate Empirical Mode Decomposition (MEMD)~\citep{rehmanMultivariateEmpiricalMode2010}; local extrema can not be applied to multidimensional data~\citep{rehmanMultivariateEmpiricalMode2010}.
In MEMD, multiple projections of the multivariate signal are generated along different directions in n-dimensional spaces; the multidimensional envelope of the signal is then obtained by integrating across the different envelopes over all these projections~\citep{rehmanMultivariateEmpiricalMode2010, rillingBivariateEmpiricalMode2007}.
This method yields the same number of oscillatory modes (IMFs) across the different dimensions of the multivariate signal. Also, each oscillatory mode across dimensions corresponds to the same narrow-band mode (frequency) (mode-alignment)~\citep{rehmanMultivariateEmpiricalMode2010}.

\subsection{Representation of marginal Hilbert spectrum for IMF signals\label{suppl:HilbertSpectrum_theory}}

A suitable representation of the energy or power of a non-stationary and/or non-linear signal across the full range of frequency designated as marginal Hilbert spectrum and can be obtained using Hilbert Spectral Analysis. This should not be misinterpreted as Power Spectral Density (PSD) representation of a stationary signal extracted using Fourier Transform. In a PSD, peak values in power at certain frequencies indicate a cos or sin wave to be pronounced and contribute to the signal throughout the whole time span. However, pronounced values of energy or power within the marginal Hilbert spectrum indicate that there is a greater likelihood to be a pronounced sin or cos component prevalent within the signal spread locally in time \citep{huangEmpiricalModeDecomposition1998}. In order to calculate the marginal Hilbert spectrum, firstly we apply the Hilbert transform \citep{huangEmpiricalModeDecomposition1998, huangConfidenceLimitEmpirical2003, huangHilbertHuangTransformIts2014} to each univariate signal (dimension of IMF cycles) for deriving their time-varying characteristics (instantaneous frequency, phase, and amplitude).

For any (real-valued) univariate signal $u(t)$, we can derive its Hilbert transform as:

\begin{equation}
    H(u)(t) = \frac{1}{\pi}P\int_{-\infty}^{+\infty} \frac{u(\tau)}{t-\tau}d\tau,
\end{equation}
where $P$ represents the Cauchy principal value for any function $u(t) \in L^{P}$ class~\citep{huangEmpiricalModeDecomposition1998}. 

The analytical signal $v(t)$ obtained from the Hilbert transform can be expressed as:
\begin{equation}
    v(t) = u(t) + iH(u)(t) = a(t)e^{i\theta(t)},
\end{equation}

where 
\begin{equation}
a(t) = \sqrt{u(t)^2 + H(u)(t)^2}
\end{equation}

and 
\begin{equation}
\theta(t) = \tan^{-1}\left( \frac{H(u)(t)}{u(t)} \right)
\end{equation}

where $a(t)$ and $\theta(t)$ are the instantaneous amplitude and instantaneous phase, respectively.

The instantaneous frequency, $f(t)$, can then be calculated as follows:
\begin{equation}
    f(t) = \frac{d\theta(t)}{dt}.
\end{equation}

Through the Hilbert spectral analysis, each IMF's instantaneous frequency can be represented as functions of time. Thus, an energy-frequency-time distribution can be obtained for each IMF signal named as the Hilbert energy spectrum or Hilbert spectrum.

For each univariate IMF signal, we can obtain the Hilbert spectrum (the squared value of the amplitude) as a function of instantaneous frequency and time using the following equation:

\begin{equation}
    H(f,t)= \left\{
        \begin{array}{ll}
        a^{2}(t), &  f = f(t)\\
        0, & \text{otherwise.}\\
    \end{array} \right. 
    \label{eq:Hi(f,t)}
\end{equation}

For better clarity through visualisations, we will display the inverse of the instantaneous frequency, i.e. the instantaneous cycle period. %also termed "cycle length" throughout this manuscript.

Taking into consideration a well-defined Hilbert spectrum $H(f,t)$, we can then obtain the marginal Hilbert spectrum $h(f)$ of the signal $u(t)$. $h(f)$ is the total energy distributed across the frequency space within a time span of $[0,T]$. Mathematically, this is defined as:

\begin{equation}
    h(f) = \int_{0}^{T}H(f,t)dt.
    \label{eq:h(f)}
\end{equation}

In order to obtain a Hilbert spectrum representation for multivariate signals based on equation \ref{eq:Hi(f,t)}, we simply averaged over the dimensions $H_{i}(f,t)$ across $i = 1, \dots, k$ dimensions: 

\begin{equation}
    \Bar{H}(f,t) = \frac{\sum_{i=1}^k H_i(f,t)}{k}.
    \label{eq:h1(f)}
\end{equation}

Then, equation \ref{eq:h(f)} is applicable for obtaining the marginal Hilbert spectrum for a univariate IMF signal. Because in our case we have multivariate IMF signals we will compute the marginal Hilbert spectrum of each multivariate IMF signal across all dimensions. In order to do that we are going to integrate over the full time span the averaged Hilbert spectrum across all dimensions, $\Bar{H}(f,t)$.

This can be expressed as:

\begin{equation}
    \Bar{h}(f) = \int_{0}^{T}\Bar{H}(f,t)dt.
    \label{eq:h2(f)}
\end{equation}

For numerical computations, we discretised time $t$ to compute the integrals as sums. 
For each multivariate IMF signal, a marginal Hilbert (energy) spectrum was obtained. For defining which frequency of the initial signal is representative from each IMF, we identified the dominant frequency as the frequency with the global maximum energy value within the marginal Hilbert (energy) spectrum representation. Then, each IMF signal is characterised by a pair of (energy/power, frequency) and this can be visualised for each subject as can be seen in Supplementary Figures \ref{fig:char_ampl_freq1} and \ref{fig:char_ampl_freq2}.

\section{Choice of band power cycles for further analysis\label{suppl:choice_of_cycles}}

\begin{figure}[h!]
    % \vspace{-1cm}
    % \hspace{-1cm}
    \centering
    \includegraphics[scale = 0.9]{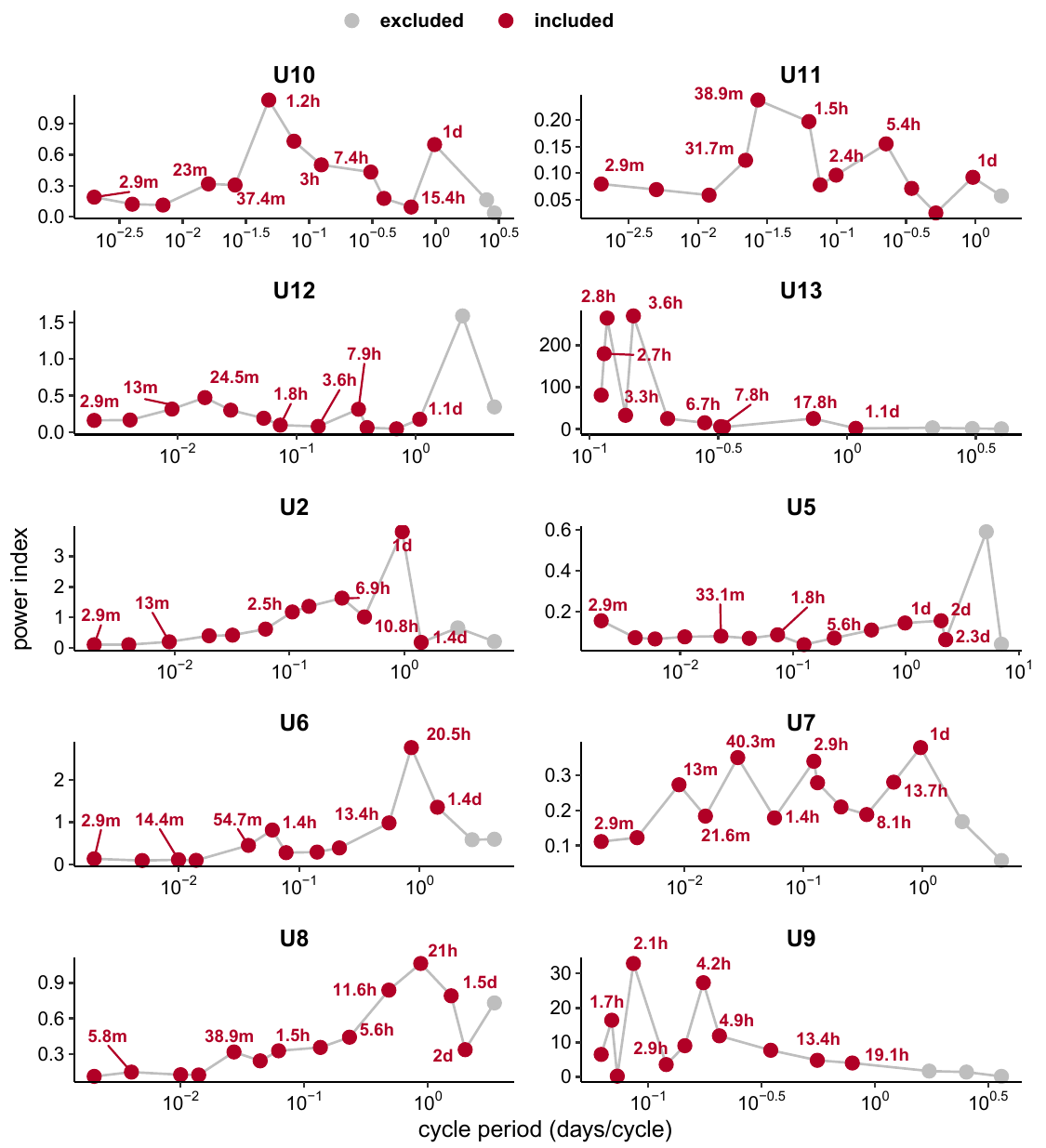}
       \caption{\textbf{Characteristic cycle period and energy for the band power IMF cycles as obtained from MEMD}. Each plot corresponds to each subject and represents the power for each cycle period (days/cycle) of all band power IMF cycles. Red coloured circles indicate cycles that were included the analysis, while gray coloured ones are the ones excluded. The band power IMF1 cycle as well as the band power residue are excluded from each subject's plot. Only selected labels shown in the graph for better clarity and visualisation.}
    \label{fig:char_ampl_freq1}
\end{figure}

\begin{figure}[h!]
    % \vspace{-1cm}
    % \hspace{-1cm}
    \centering
    \includegraphics[scale = 0.9]{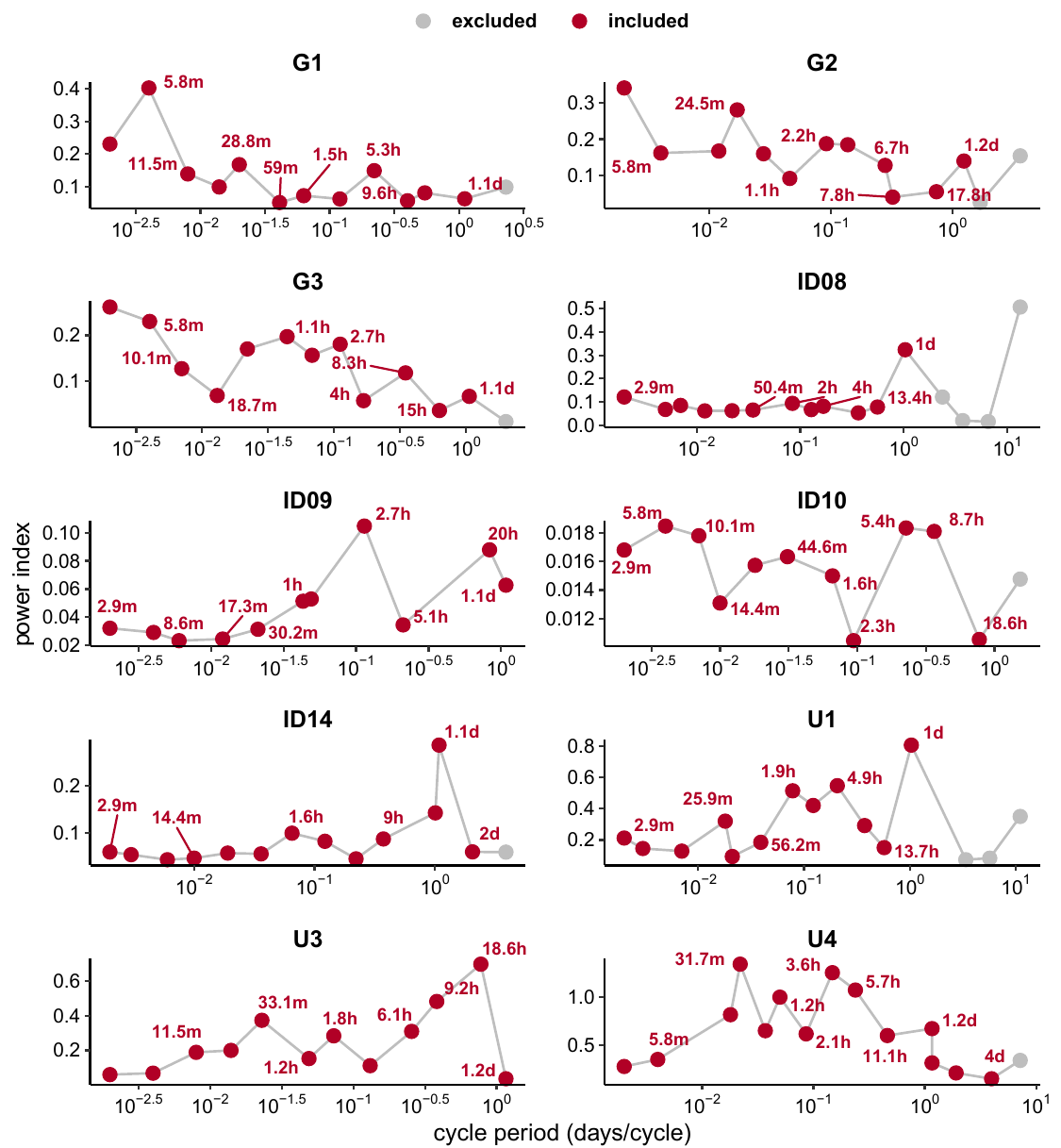}
       \caption{\textbf{Characteristic cycle period and energy for the band power IMF cycles as obtained from MEMD}. Equivalent figure to Fig.~\ref{fig:char_ampl_freq1}}
    \label{fig:char_ampl_freq2}
\end{figure}

In order to select appropriate IMF signals for further analysis, we used the cycle period (inverse of frequency) along with the power for each IMF (Supplementary Figures \ref{fig:char_ampl_freq1} and \ref{fig:char_ampl_freq2}). Cycles lower or equal to the circadian cycle if apparent (cycle period $\leq$ 1 day/cycle) were included in the analysis. Cycles greater than 1 day/cycle were included in the analysis, if their cycle period was at least $\frac{1}{3}$ of the total recording indicating that this cycle is more stable and we are more confident that this cycle is embedded within the signal as we do not rely on observing one cycle only, but at least three cycles within the full recording. Thus, in most of the cases the last three or more slowest band power cycles including the band power residue were extracted from the overall analysis. Finally, the fastest band power cycle (named IMF1) was not included in the analysis as it often captures noise from the raw signal.

\newpage

\newpage

\section{Linear-Circular correlation between the phase of band power cycles and severity measure -- a univariate measure \label{suppl:CircLinearCorr}}

For each patient, we compared the seizure duration for every seizure to the patient's band power cycles; for each mode within each frequency band (dimension), we found the mode phases of seizures based on the associated onset times and further correlated those with seizure duration.
Mode phases are angular data and thus common correlation metrics, such as Pearson or Spearman's rank correlation coefficients are not appropriate. To determine the association between seizure duration (linear variable) and mode phase in a frequency band (circular variable) we computed Mardia's non-parametric Linear-Circular Correlation coefficient which measures association using ranks \citep{mardiaLinearCircularCorrelationCoefficients1976}. This is a non-parametric measure, which means that it does not assume any form of underlying distribution for both the linear and circular variables.
It can be thought of as the equivalent to Spearman’s rank correlation coefficient between a linear and a circular random variable.

\subsubsection{Mardia's Rank Correlation Coefficient\label{suppl:MardiasCorr}}

A measure of association between two random variables $X$ and $\Phi$, where $X$ is a linear variable on $(-\infty, +\infty)$ and $\Phi$ a circular variable on $[0, 2\pi)$ was proposed by Mardia \citep{mardiaLinearCircularCorrelationCoefficients1976} where the random sample, $(x_{i}, \phi_{i}), i = 1, 2, \dots, n$, of observations on $(X, \Phi)$ are re-arranged by their ranks; $x_{i}$ and $\phi_{i}$ are arranged by the ranks $i$ and $r_{i}$, respectively. The observations of the linear variable are first reordered from smallest to largest, so as $x_{1} \leq \dots \leq x_{n}$ and assigned a corresponding rank, $i$, based on the associated index $i = 1, \dots, n$. For the circular observations $\phi_{i}$, then a circular ranked variable $m_{i}$ can be defined as:
\begin{equation}
    m_{i} = \frac{2\pi (r_{i})}{n},
\end{equation}
where $\{r_{i}, i = 1, \dots, n\}$ represent the circular ranks and $n$ is the sample size of the linear-circular pairs of observations, $(x_{i}, \phi_{i})$.
Then, Mardia's rank correlation coefficient is defined as:
\begin{equation}
    U_{n} = \frac{24(T_{C}^2 + T_{S}^2)}{n^2(n+1)} \sim X_{2}^2 \text{ for } n \xrightarrow[]{} \infty,
\end{equation}

where $X_{2}^2$ is the $X^2$ distribution with 2 degrees of freedom and 
\begin{equation} \label{eq1}
\begin{split}
T_{C} & = \sum_{i=1}^{n} {x_{i}cos(m_{i})} \\
T_{S} & = \sum_{i=1}^{n} {x_{i}sin(m_{i})}.
\end{split}
\end{equation}
Higher values of $U_{n}$ indicate stronger association. Also, $U_{n}$ is invariant under a change of origin for the linear variable, $X$ or under rotations of the circular variable, $\Phi$.

However, for reporting the correlation we used a scaled measure of the $U_{n}$ termed $D_{n}$. This is because the $U_{n}$ does not scale within $[0,1]$, as the common $R^2$ values. Thus, using a suitable transformation function, $a_{n}$, we can obtain a scaled version of the Mardia's rank correlation coefficient within the range $[0,1]$ as follows:

\begin{equation}
    D_{n} = a_{n}(T^2_{c} + T^2_{S}),
\end{equation}

where if $n$ is even,  

\begin{equation}
    a_{n} = \frac{1}{1+5cot^{2}(\frac{\pi}{n}) + 4cot^{4}(\frac{\pi}{n})},
\end{equation}

whereas if $n$ is odd,

\begin{equation}
    a_{n} = \frac{2sin^{4}(\frac{\pi}{n})}{(1 + cos(\frac{\pi}{n}))^{3}}.
\end{equation}

\subsubsection{Randomisation test \label{suppl:randomisationTestCorr}}

For large $n$, $U_{n}$ under independence follows approximately the $X^{2}$ on two degrees of freedom if $X$ and $\Phi$ have continuous distributions.
Significance of the independence can be established using the $U^{*}_{n} = (T^{2}_{C} + T^{2}_{S})$ measure of association \citep{pandolfoCircularStatistics2015} based on a randomisation test. We use $U^{*}_{n}$ as it is less computationally intensive.
For the randomisation test, we randomly obtain the circular ranking and further define the circular ranked variable $m_{i}$ which is paired randomly with the linear ranks $i$. The $U^{*}_{n}$ measure was obtained for $1,000$ random selected linear and circular ranking to form a null distribution. The p-value of the observed $U^{*}_{n}$ measure was the percentage of times a $U^{*}_{n}$ measure of association obtained from the random selected ranking was greater or equal to the observed correlation coefficient, $U^{*}_{n}$. 

\subsubsection{Linear-circular correlation between seizure duration and the phase of band power cycles}

\begin{figure}[h!]
    %\hspace{-0.2cm}
    \centering
    \includegraphics[scale = 0.9]{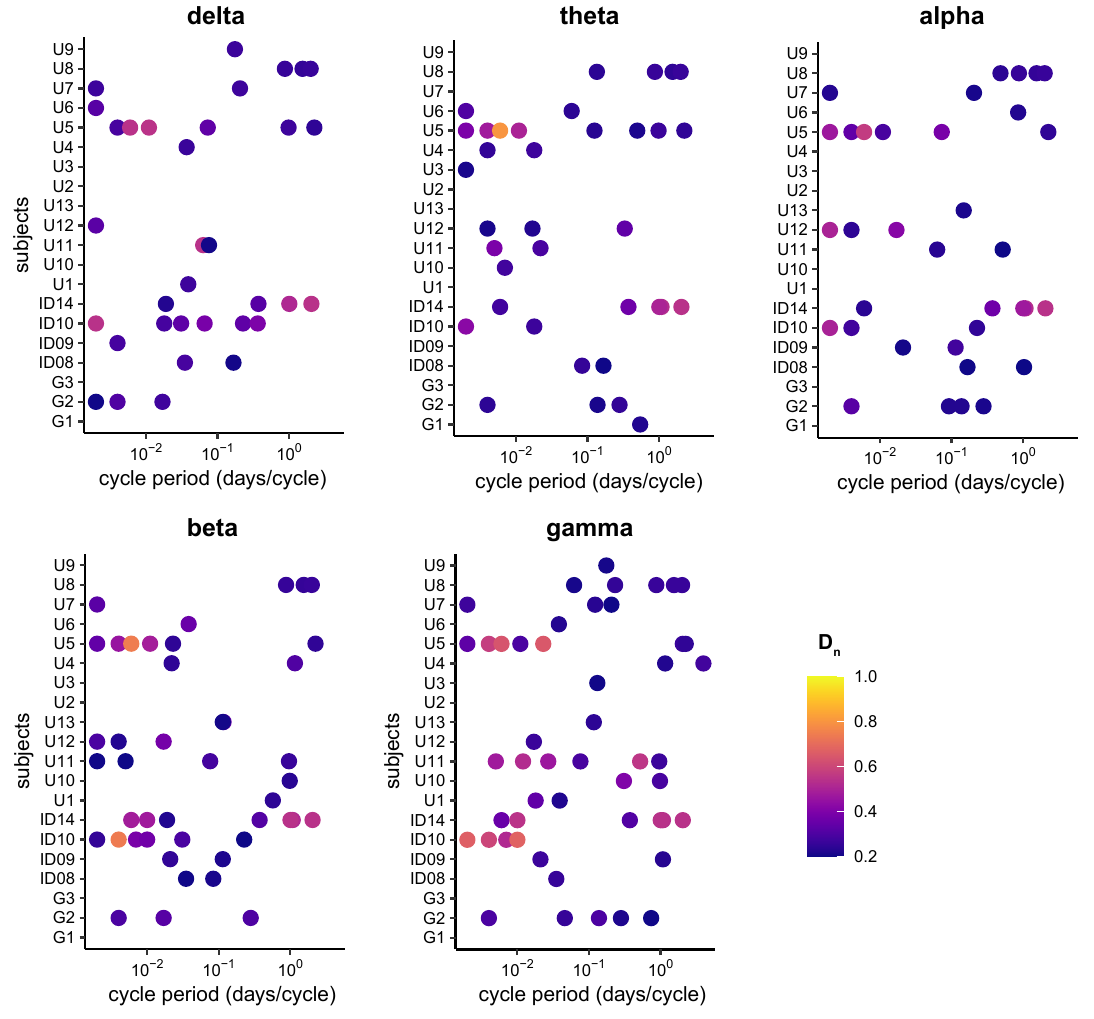}
       \caption{\textbf{Associations of seizure duration with phases of band power cycles}. Dot plots of the Mardia's rank correlation between the seizure duration and the phases of the band power cycles across all subjects. Each dot plot represents one main frequency band. Band power cycles from weak associations ($D_{n} \leq 0.2$) are not shown for clarity of visualisation.}
    \label{fig:Mardias_corr_all_freq_bands}
\end{figure}

To determine the association between the seizure duration and the phase of each band power cycle at which the seizure occurred we performed the rank linear-circular correlation $D_{n}$ \citep{mardiaLinearCircularCorrelationCoefficients1976} for each subject. Supplementary Fig.~\ref{fig:Mardias_corr_all_freq_bands} shows the correlations between the seizure duration and the phases of band power cycles in each frequency band across subjects. Overall, there are some weak to moderate correlations across all frequency bands.

\section{Modeling seizure duration using linear-circular regression}

\subsection{Linear-circular regression for cylindrical data\label{suppl:CircRegression}}

For modelling the relationship between a linear response variable, $Y_{t}$, and a circular random variable $\phi_{t}$, $t = 1, 2, \dots, n$ we used the following cosine regression \citep{mardiaLinearCircularCorrelationCoefficients1976, pewseyRecentAdvancesDirectional2020}:

\begin{equation}
    Y_{t} = \mu + \beta cos(\phi_{t}- \phi_{0}) + \epsilon_{t}
\end{equation}

The $n$ represents the total number of seizures, $\phi_{.}$ is the instantaneous phase for the specific IMF cycle, $Y_{.}$ is the linear response of interest (here, the log seizure duration) and $\phi_{0}$ is the so-called acrophase angle. This is the phase of the corresponding IMF cycle where seizure occurrence reached its peak in the corresponding IMF cycle, as the sample of phases for every IMF cycle was chosen based on the onset times.  

The above cosine formula can be rewritten as:

\begin{equation}
    Y_{t} = \mu +  \gamma cos(\phi_{t}) + \delta sin(\phi_{t}) + \epsilon_{t}  \label{lincircreg}
\end{equation}

This regression can be thought of as a multiple linear regression of $Y$ on ($cos(\phi_{t}), sin(\phi_{t})$). 

Extending this regression from the univariate case to the multivariate case by using multiple ($K$) IMF cycles, this can then be written as \citep{mardiaDirectionalStatistics1999}:

\begin{equation}
    Y_{t} = \mu + \sum\limits_{k=1}^K \beta_{k}cos(\phi_{t}^k- \phi_{0}^k) + \epsilon_{t}
\end{equation}

Finally, the above cosine equation can be rewritten as:

\begin{equation}
    Y_{t} = \mu + \sum\limits_{k=1}^K \gamma_{k}cos(\phi_{t}^k) + \sum\limits_{k=1}^K\delta_{k}sin(\phi_{t}^k) + \epsilon_{t}  \label{lincircreg}
\end{equation}

In all the different forms of the cosine model, $\mu$ is the expected mean value of the linear response variable $Y$. $\beta_{k} = \sqrt{\gamma_{k}^{2} + \delta_{k}^{2}}$ is the magnitude of contribution of the $k$-th IMF cycle.

Equation \ref{lincircreg} can be fitted based on data using a multiple regression framework yielding estimates for $\gamma_{k}$, and $\delta_{k}$, for each $k$-th IMF cycle. Subsequently, we can thus obtain the magnitude of contribution by each $k$-th IMF through $\beta_{k}$. Previous study \citep{proixForecastingSeizureRisk2021} has used a similar approach for including cycles of multiple timescales into a linear generalised regression model. Proix et al. obtained cycles from interictal spike rate data using wavelet analysis, in order to predict whether a seizure will occur or not. So the response variable was binary as opposed to our study, where the response can take values in $\mathbb{R}$.

\subsection{Forming models for seizure duration\label{suppl:formingModels}}

\begin{figure}[h!]
    \hspace{-0.2cm}
    % \centering
    \includegraphics[scale = 1]{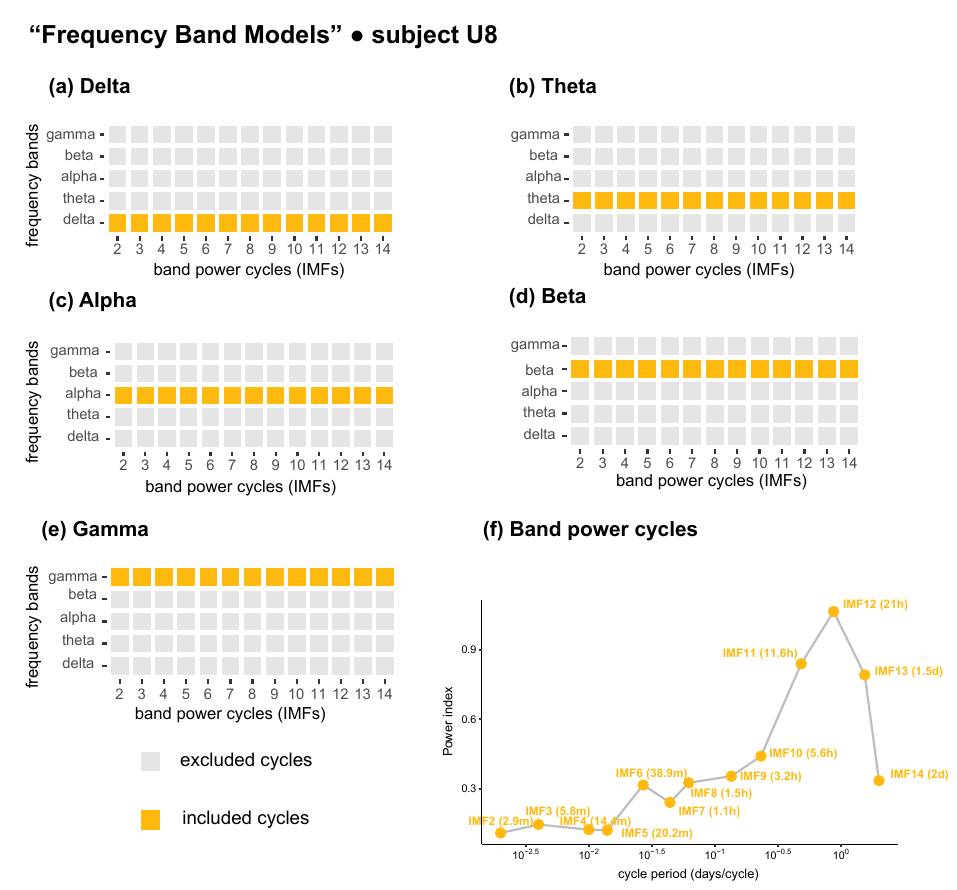}
      \caption{\textbf{Illustration of "frequency band" models in an example subject}. (a-e) Heatmaps of all the different band power cycles for all frequency bands. The rows depict the frequency bands and the columns the spectrum of the narrow-band band power cycles selected for further analysis (see Suppl. section~\ref{suppl:choice_of_cycles}). Cells coloured with yellow indicate the corresponding variables used in each "frequency band" model. (f) Characteristic power and cycle period (days/cycle) of cycles included in each of the frequency band models with text displaying the band power cycle (IMF) and the characteristic cycle period in parenthesis.}
    \label{fig:frebands_models}
\end{figure}

\begin{figure}[h!]
    \hspace{-0.2cm}
    % \centering
    \includegraphics[scale =1]{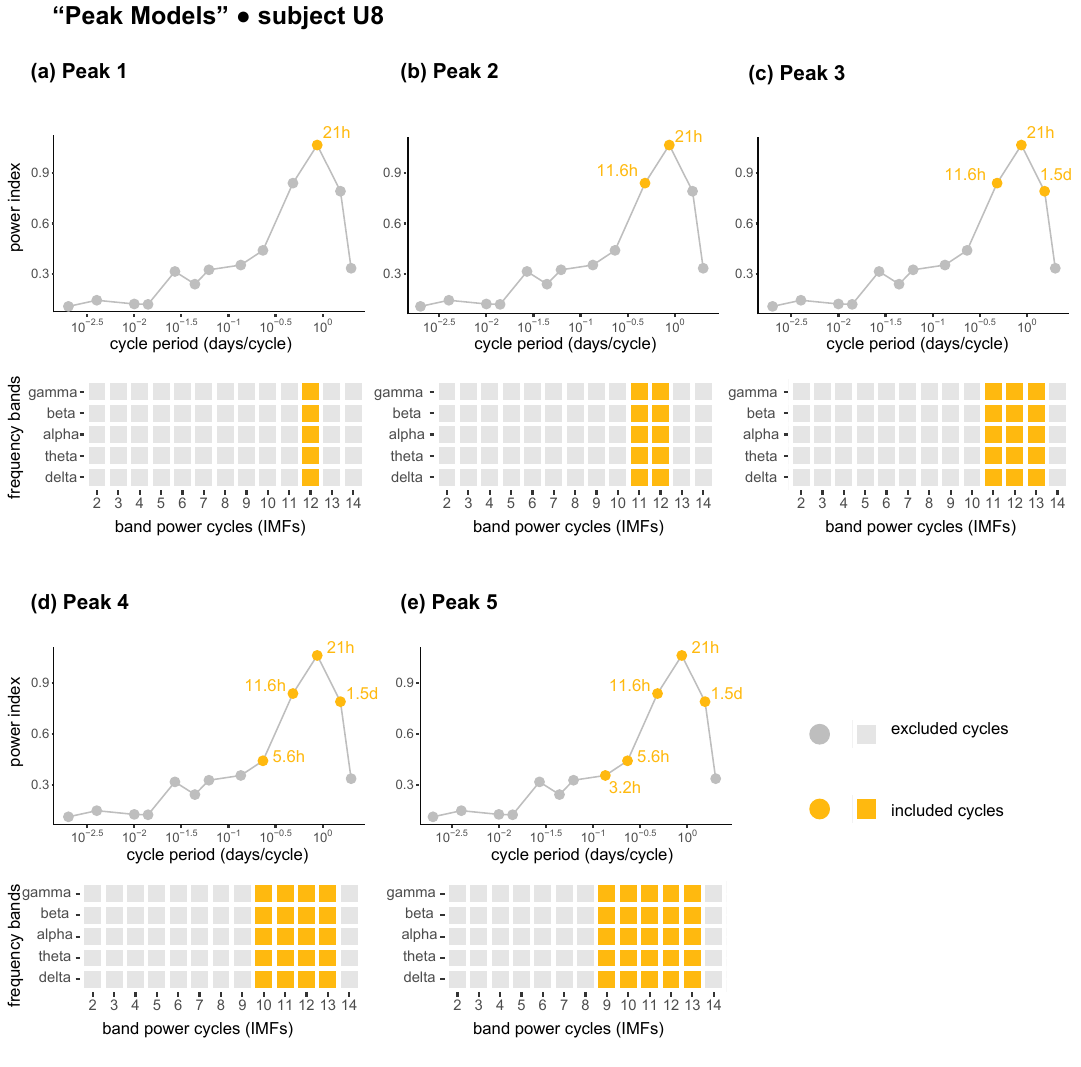}
      \caption{\textbf{Illustration of "Peak" models in an example subject}. (a-e) Within each panel, the top plots show the characteristic power and cycle period (days/cycle) of all band power cycles selected for further analysis (see Supplementary section~\ref{suppl:choice_of_cycles}) represented by circle dots for an example subject. Circle dots coloured in yellow depict the band power cycles included in the corresponding "peak" models with text displaying their characteristic cycle period. The bottom plots are a visual representation of all band power cycles across all frequency bands that share the same cycle period. Variables included in the corresponding "peak" models are coloured with yellow.}
    \label{fig:peak_models}
\end{figure}

We wanted to investigate the relationship of seizure duration with combinations of band power IMFs extracted from the MEMD. We fitted 12 models per subject, each with a different selection of explanatory variables. The 12 models can be grouped into four categories, the "frequency band models" (five models) (see Fig.\ref{fig:frebands_models}), the "peak models" (five models) (see Fig.\ref{fig:peak_models}), the full (all explanatory variables) and the intercept (without any explanatory variables) models. 

The band power cycles from a given frequency band formed the explanatory variables for the corresponding "frequency band model" (see Fig.~\ref{fig:frebands_models}). Note that all cycles corresponding to the cycle periods that selected for further analysis (see Supplementary section \ref{suppl:choice_of_cycles}) were identified within each frequency band and included in the corresponding "frequency band model". Fig. \ref{fig:frebands_models}f illustrates the characteristic power and cycle period of cycles included in each of the "frequency band models" for an example subject, U8. For each one of the five "peak models", we used the band power cycles across all frequency bands that appeared to have more prominent amplitudes (see Supp. Figures \ref{fig:char_ampl_freq1} and \ref{fig:char_ampl_freq2}). Starting from the "peak 1" model that contains the band power cycles (across all frequency bands) with the highest power, we then constructed the "peak 2" model by adding the two most prominent band power cycles across all frequency bands, and sequentially formed the remaining "peak models" (Fig.~\ref{fig:peak_models}). Finally, we formed the full model and the intercept one. The former included all the band power cycles across all frequency bands as independent variables, while the latter had no variables, just the intercept term.

%For every subject, we used a linear-circular regression framework (see Section \ref{subsec:CircLinearModel}), where each model included the seizure duration as response variable and the corresponding band power cycles' phases based on the evaluated model.  

Initially, we performed a variable selection step for our analysis, as the number of explanatory variables was relatively large compared to the sample size for all models except the intercept one. We used group-LASSO \citep{yuanModelSelectionEstimation2006, brehenyGroupDescentAlgorithms2015}, which is a sparse shrinkage method and formed groups of the $sin()$ and $cos()$ terms for each band power cycle involved in the model. For the LASSO, the tuning parameter $\lambda$ was selected using a 10-fold cross validation method from a range of values $\lambda = 10^{-3}, 10^{-2.92}\dots, 10^{4.92}, 10^{5}$. 

After selecting a small number of explanatory variables, a linear-circular regression was performed for each subject. We obtained adjusted $R^2$ as a model performance metric. A leave-one-out cross-validation (LOOCV) was also performed as an additional metric for assessing model performance. Both performance metrics were used to select the "best" model. The adjusted $R^2$ was used to assess how well the model can explain the variability observed in seizure duration, while the LOOCV was taken into account for accurately estimating the out-of-sample error (overfitting) \citep{burmanComparativeStudyOrdinary1989, arlotSurveyCrossvalidationProcedures2010}.
% In the case of linear regression, LOOCV provides the
% least biased and lowest variance estimate of out-of-sample
% error among other CV methods [Burman, 1989]

\begin{figure}
\centering
\includegraphics[scale = 0.9]{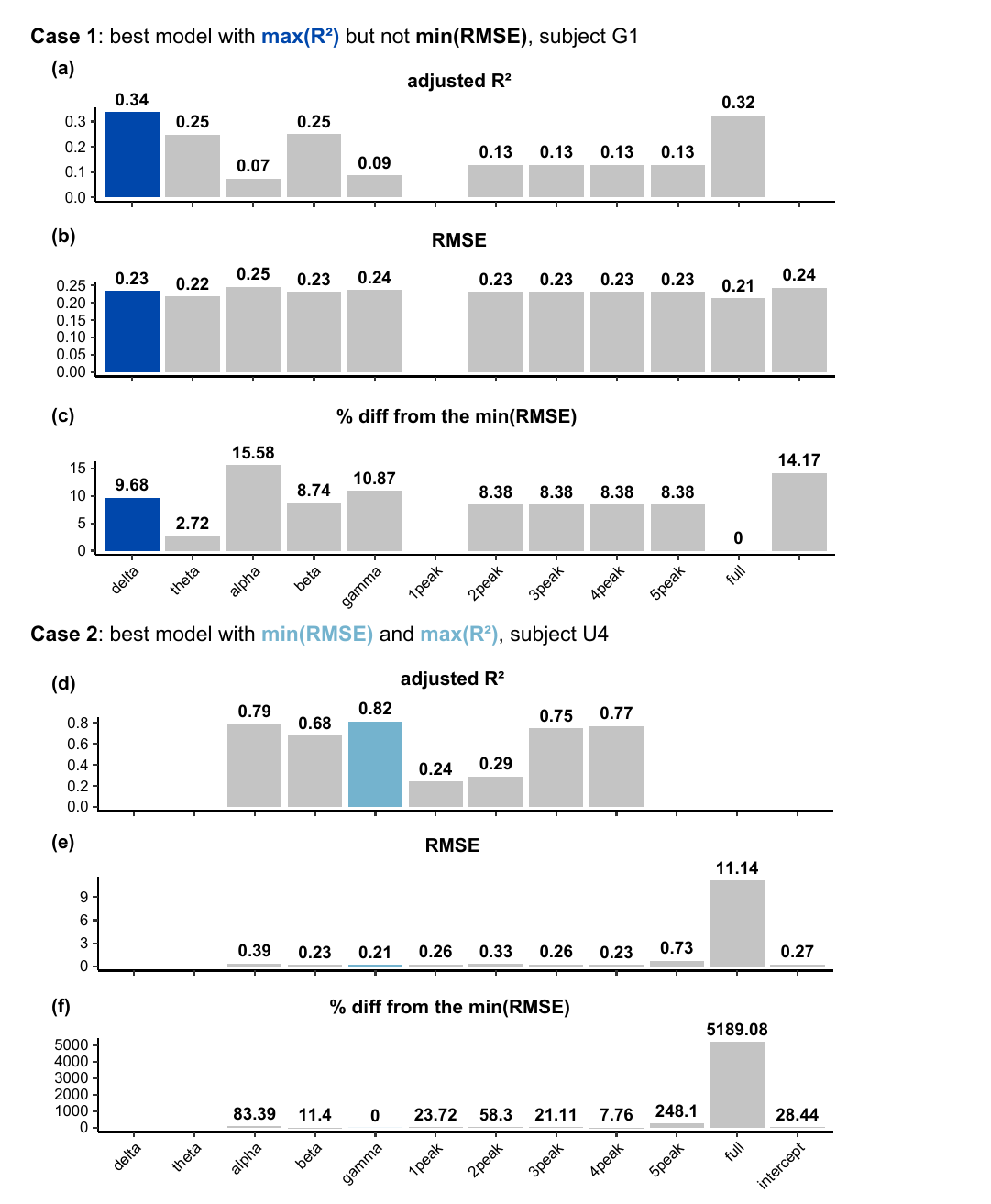}
\caption{\textbf{Selection of the "best" model in two example subjects}. (a-c) Representation of model selection in an example subject, where the model with min(RMSE) obtained from the LOOCV did not correspond to the model with $\text{max}(R^2)$. (d-f) Illustration of model selection in an example subject, where the "best" model had min(RMSE) obtained from the LOOCV and $\text{max}(R^2)$. (a,d) Bar plots of the adjusted $R^2$ obtained across all models for two example subjects. Non-grey coloured bars indicate the final selected model. (b,e) Bar plots of the RMSE obtained from the LOOCV. (c,f) Bar plots representing the \%absolute difference between the RMSE of each model with the min(RMSE).}\label{fig:model_selection}
\end{figure}

For choosing the best model, we first identified the model with $RMSE_{0}=\text{min}(RMSE)$, as well as the one with $\text{max(adjusted}~R^2)$. If these two models matched, then the final selected model was determined (see Supplementary Fig.~\ref{fig:model_selection} Case 2). In cases where the model with $\text{min}(RMSE)$ did not have the $\text{max(adjusted}~R^2)$ (see Supplementary Fig.~\ref{fig:model_selection} Case 1), we computed the absolute percentage difference of the RMSE corresponding to all models with the $RMSE_{0}$, $RMSE_{diff}=abs(RMSE-RMSE_{0})/RMSE_{0})*100\%$ (see Supplementary Fig.~\ref{fig:model_selection} c and f). If there was a model with higher adjusted $R^2$ and $\%RMSE_{diff}< 20\%$, then this was selected as "best". In cases where the $RMSE_{0}$ corresponded to the "intercept" model, then this was chosen as "best" model (this occurred only in one subject, U12).

\newpage

\subsection{Performance of the "best" model based on its predictive accuracy \label{suppl:performanc_predict}}

\begin{figure}
\centering
\includegraphics[scale = 0.85]{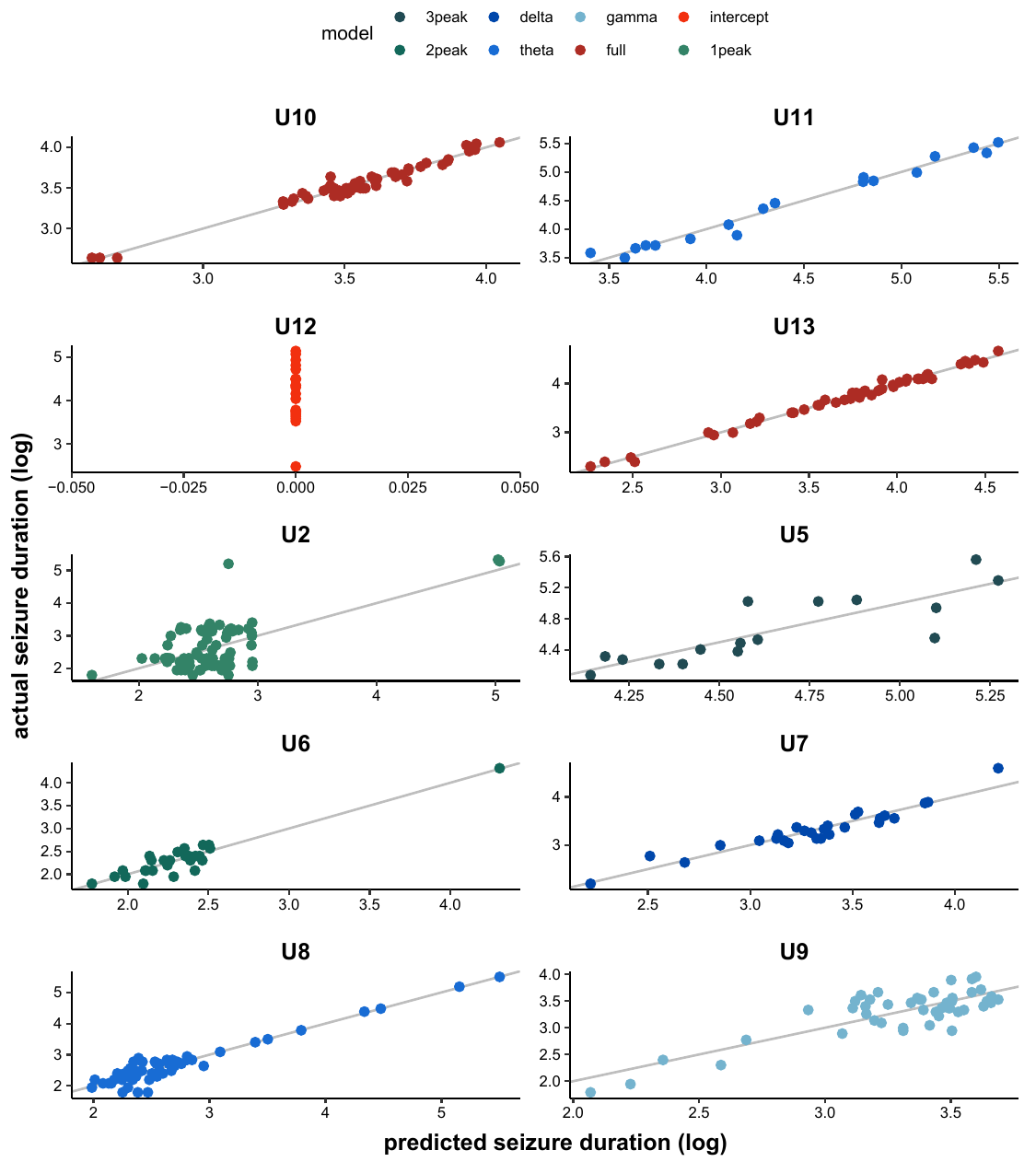}
\caption{\textbf{Actual against fitted values across all subjects}. Scatterplots of the actual against the predicted values for seizure duration (log scale) for each subject as obtained from the final "best" models.}\label{fig:actual_vs_fitted_all1}
\end{figure}

\begin{figure}
\centering
\includegraphics[scale = 0.9]{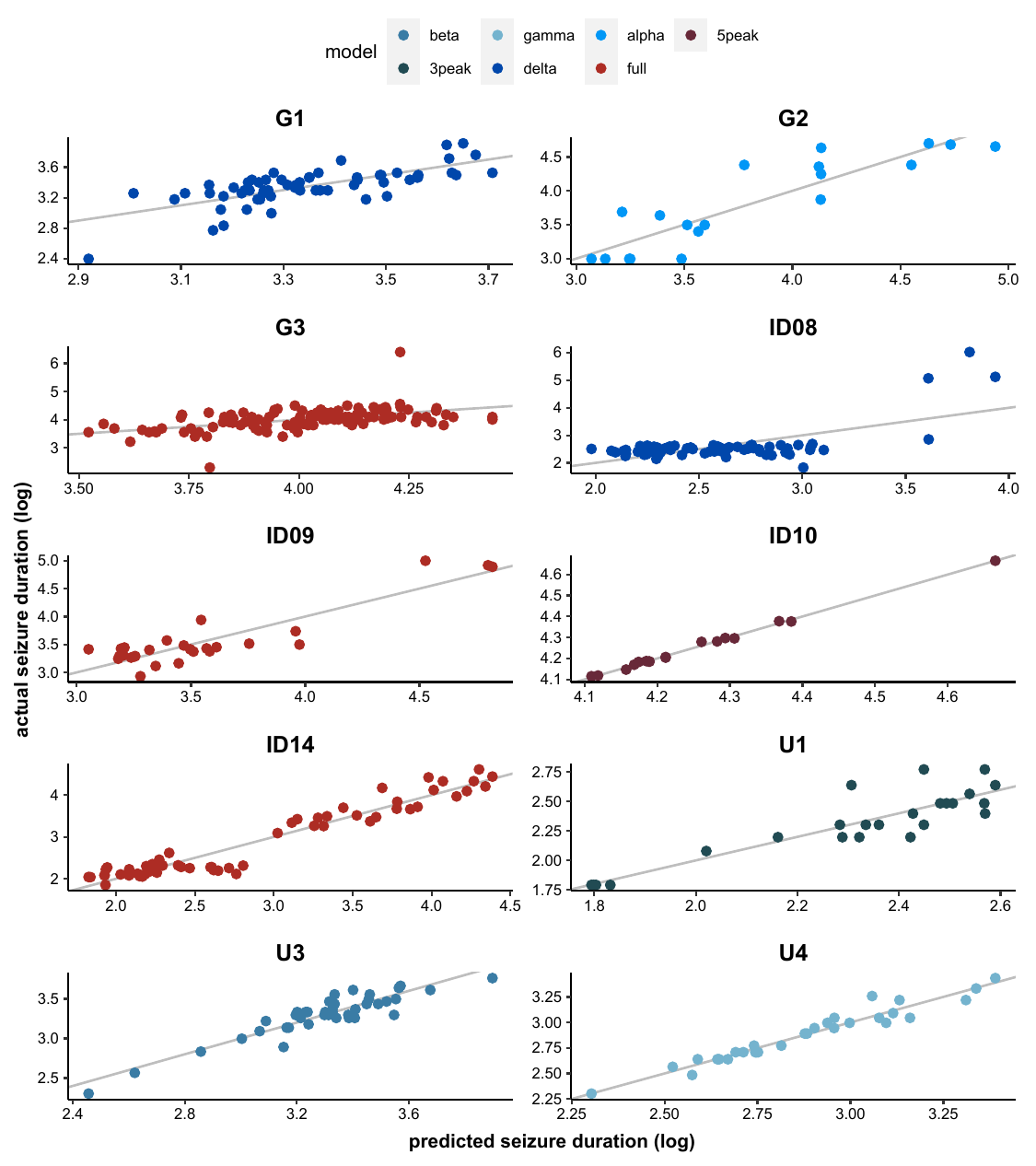}
\caption{\textbf{Actual against fitted values across all subjects}. Equivalent to Fig. \ref{fig:actual_vs_fitted_all1}.}\label{fig:actual_vs_fitted_all2}
\end{figure}

For each subject, we performed a circular-linear model selection and regression (see Supplementary Section \ref{suppl:CircRegression}). The performance of each subject's "best" model can be seen in Suppl. Figures \ref{fig:actual_vs_fitted_all1} and \ref{fig:actual_vs_fitted_all2}, where the predicted seizure durations based on band power cycles were plotted against the actual seizure duration values.

\newpage

\subsection{Performance of the "best" models based on permutation tests \label{suppl:permutation_performance}}

\begin{figure}
\centering
\includegraphics[scale = 0.9]{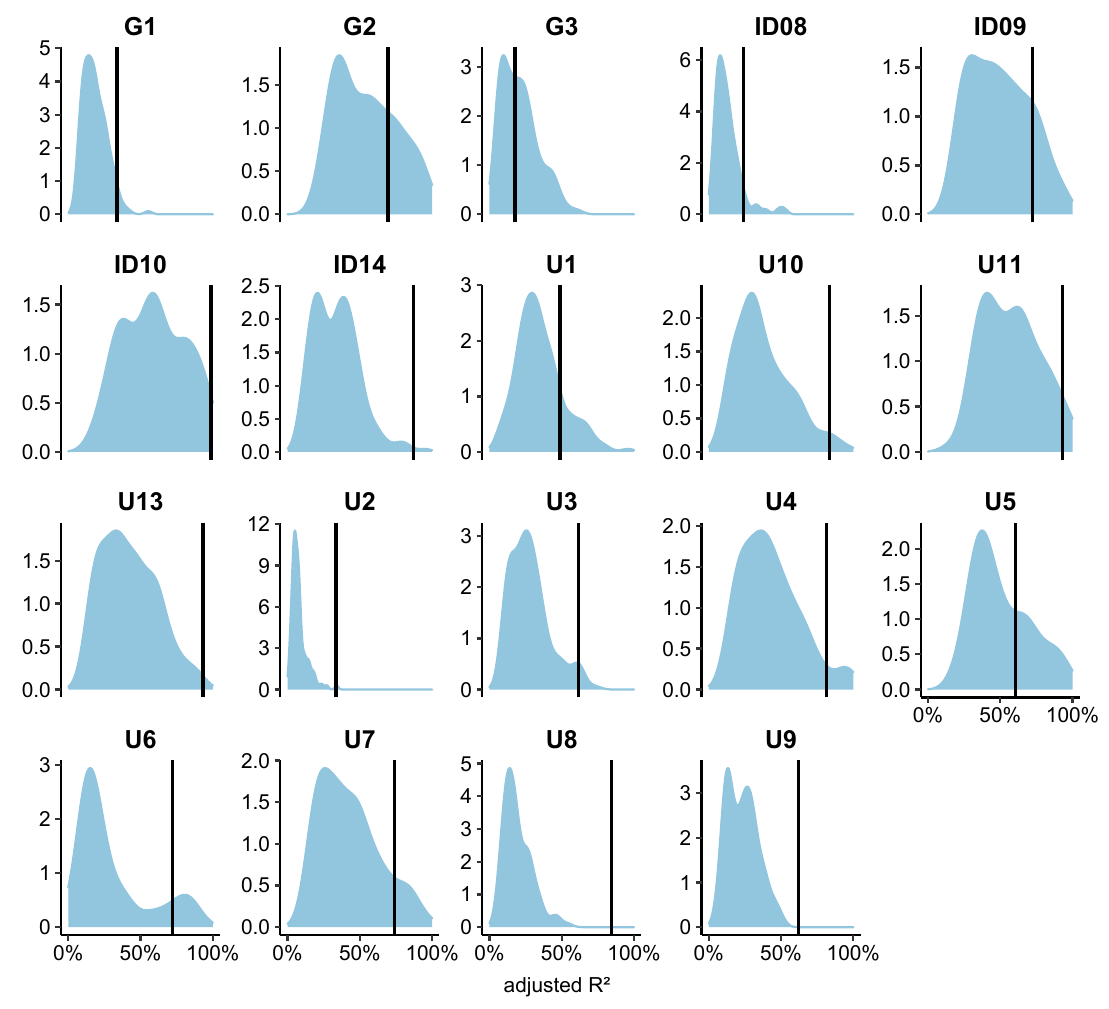}
\caption{\textbf{Distribution of the adjusted $R^2$ values using permuted seizure orders.} The seizure duration was used as response variable, while the band power cycles were included in the model as explanatory variables. The vertical black line represents the adjusted $R^2$ for the same analysis performed on the original seizure order (see Suppl. Sections~\ref{suppl:CircRegression} and \ref{suppl:formingModels}).}\label{fig:perm_A}
\end{figure}

To assess the performance of the "best" models we performed a permutation test. In each permutation iteration, we first randomly permuted the order of the seizures by randomly shuffling the response variable (seizure duration). Then, using the band power cycles from the selected model, we applied group-LASSO and circular-linear regression including only those variables (keeping them unchanged). Finally, we calculated the adjusted $R^2$ for each iteration.  We repeated this procedure as many times as the number of permutations, which we selected as 500. Statistical significance was determined based on a significance level of $5\%$.

% Similarly to the analysis described in the previous section, we additionally performed a permutation test. In each permutation iteration, we first randomly permuted the order of the seizures (but not their timing). Then, we obtained new IMF seizure distance matrices and performed the LASSO and linear regression, exactly as described in the Section~\ref{subsec:Linear analysis}, leaving the response variable unchanged. Finally, we calculated the adjusted $R^2$ for each iteration (see Fig.~\ref{fig:adjusted_Rsquared_shuffle}). Statistical significance was determined based on a significance level of $5\%$. Again, the aforementioned steps were performed for all subjects with at least six recorded seizures. Significance levels were similarly for all tested subjects as in the previous section.
% \begin{figure}[H!]
%     \vspace{-2cm}
%     % \centering
%     \includegraphics[scale = 1]{figures/model_performance_sz_dur_log_perm.pdf}
%       \caption{\textbf{Permutation results across subjects}. }
%     \label{fig:perm_B}
% \end{figure}

% \begin{figure}[H!]
%     %\hspace{-0.2cm}
%     \centering
%     \includegraphics[scale =1]{figures/Summary_models_sz_dur_RMSE.pdf}
%       \caption{\textbf{"Best" models across subjects}. 
%       (a) Barplot of the RMSE as computed from LOOCV applied to the "best" model selected for each subject. Red asterisks represent the significant models as those determined by a permutation test for the adjusted $R^2$ (see Supplementary section \ref{fig:perm_A}).}
%     \label{fig:all_models_RMSE}
% \end{figure}

\subsection{Correction for multiple comparisons\label{suppl:FDR}}

We adjusted all p-values using the Benjamini-Hochberg false discovery rate FDR correction for multiple comparisons \citep{benjaminiControllingFalseDiscovery1995}. FDR correction was applied for all statistical tests across all subjects (except permutation tests). The significance of Mardia's rank correlation was assessed using randomization tests as described in Supplementary section \ref{suppl:randomisationTestCorr} yielding uncorrected $p$-values. After applying FDR correction, the significance level was set to $5\%$.

\subsection{Statistical analysis for other severity measures}

\begin{figure}
\centering
\includegraphics[scale = 0.9]{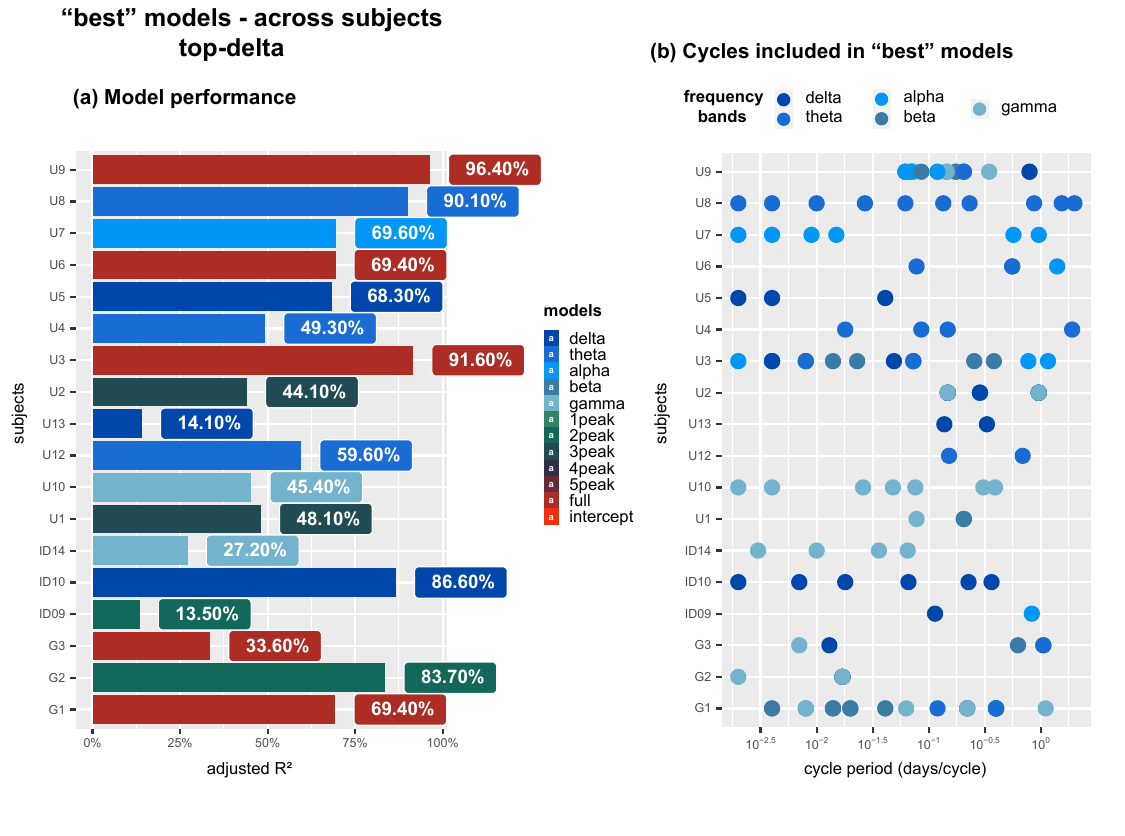}
\caption{\textbf{Selected models for one peak marker of severity across subjects}. (a) Barplot of the adjusted $R^2$ as obtained from the selected model for each subject. (b) Dot plot illustrating the cycle period of the variables selected in the final model for each subject. Colours depict the frequency band associated with these band power cycles.}\label{fig:top_delta_summary}
\end{figure}

To assess whether band power cycles were associated with other markers of severity, we performed all steps of the analysis for an example severity marker \citep{gascoigneLibraryQuantitativeMarkers2023}. Specifically, we used one of the "peak" markers, the top delta, calculated as described previously \citep{gascoigneLibraryQuantitativeMarkers2023}. Briefly, this marker encapsulates the maximum level of activity in delta band power during a seizure.
Within each subject, seizures with pre-ictal period less or equal to 120s were excluded from this analysis (Suppl. Table~\ref{suppl:table_demographic}). Suppl. Fig.~\ref{fig:top_delta_summary}a shows that 10 out of 18 subjects had an adjusted $R^2 >= 60$\%. A combination of band power cycles of different timescales could explain the variability in seizure severity, as measured from this peak marker (Suppl. Fig.~\ref{fig:top_delta_summary}b). For two subjects, the intercept model performed better compared to all the models, indicating that this subject's variability in top-delta marker was not well-predicted by band power cycles. Those subjects are not included in the Suppl. Fig.~\ref{fig:top_delta_summary}.

%if the inter-seizure interval between two seizures was less then the defined pre-ictal period (120s),  then $seizure_{a}$ was excluded.

\clearpage
\newpage

\subsection{Subject Information\label{suppl:patient_data}}

\begin{sidewaystable}
%\tiny
\footnotesize
\centering % used for centering table
% \begin{tabular}{|>{$}c<{$} *{9}{c}} % centered columns (9 columns)
\begin{tabular}{|c c c c c c c c|} % centered columns (7 columns)
\toprule
Subject & Cohort & Age(yrs) & Sex & Hemisphere (Lobe) & Recording duration & \# sz analysed for seizure duration & \# sz analysed for top-delta marker\\
\midrule
\hline
%heading
\textbf{ID08} & SWEC & - & - & - & 6d & 70 & 15\\[0.5ex] % inserting body of the table
\textbf{ID09} & SWEC & - & - & - & 2d & 27 & 27\\[0.5ex]
\textbf{ID10} & SWEC & - & - & - & 2d & 17 & 17\\[0.5ex]
\textbf{ID14} & SWEC & - & - & - & 7d & 60 & 54\\[0.5ex]
\textbf{U1} & UCLH & 35 & M & L (OP) & 7d 1h & 24 & 17\\[0.5ex]
\textbf{U2} & UCLH & 43 & F & L \shortstack{(FP} & 7d 2h & 79 & 79\\[0.5ex]
\textbf{U3} & UCLH & 28 & F & R (OP) & 1d 19h & 41 & 41\\[0.5ex]
\textbf{U4} & UCLH & 28 & M & L (T) & 12d 5h & 30 & 30\\[0.5ex]
\textbf{U5} & UCLH & 21 & F & R (T) & 8d 21h & 16 & 16\\[0.5ex]
\textbf{U6} & UCLH & 26 & M & R (F) & 6d 2h & 27 & 27\\[0.5ex]
\textbf{U7} & UCLH & 43 & F & R (TOP) & 7d 22h & 27 & 27\\[0.5ex]
\textbf{U8} & UCLH & 27 & F & L (F) & 8d & 59 & 26\\[0.5ex]
\textbf{U9} & UCLH & 39 & M & L (P) & 5d 22h & 48 & 44\\[0.5ex]
\textbf{U10} & UCLH & 28 & M & R (P) & 5d 5h & 54 & 54\\[0.5ex]
\textbf{U11} & UCLH & 24 & F & R (T) & 2d 19h & 18 & 18\\[0.5ex]
\textbf{U12} & UCLH & 26 & M & R (T) & 5d 7h & 21 & 21\\[0.5ex]
\textbf{U13} & UCLH & 32 & M & L (T) & 6d 7h & 47 & 47\\[0.5ex]
\textbf{G1} & GGC & 37 & M & L (TOP) & 3d 19h & 60 & 60\\[0.5ex]
\textbf{G2} & GGC & 35 & F &  \shortstack{L (T)} & 3d 20h & 19 & 19\\[0.5ex]
\textbf{G3} & GGC & 28 & M & L (TP) & 3d 15h & 110 & 110\\[1ex] % [1ex] adds vertical space
\bottomrule
\hline
 & & & & & &  \\
\multicolumn{6}{|c|}{\shortstack[c]{\\\\\textbf{Cohort}\\\\SWEC = Sleep-Wake-Epilepsy-Center \\ UCLH = University College London Hospital\\ GGC = NHS Greater Glasgow and Clyde center \\\\\\\\ \textbf{Sex} \\ M = male \\ F = female \\\\\\\\\\\\\\ Dash indicates no available information}}  &
\multicolumn{1}{c|}{\shortstack[c]{\textbf{Lobe}\\\\T = temporal \\ F = frontal \\ P = parietal \\ O = occipital \\ IH = interhemispheric}} & 
\multicolumn{1}{c|}{\shortstack[c]{\textbf{Hemisphere}\\\\ L = left \\ R = right \\ B = bilateral}} \\
\end{tabular}
\caption{\textbf{Demographic and Clinical Information for EMU subjects.} No metadata information for SWEC cohort. Information provided from the clinical reports for UCLH and GGC cohorts. \textbf{Subject}: Patient identifier. \textbf{Hospital}: hospital where the presurgical assessment, intracranial EEG monitoring and cortical stimulation) took place. \textbf{Age}: age, in years, at the time of the presurgical evaluation (this differs from the age of onset). \textbf{Hemisphere (Lobe)}: the hemisphere (lobe) where the onset of seizures lies on based on clinical findings. \textbf{Total recording time}: total duration of the iEEG implantation (the total recording time was computed based on the edf files available). \textbf{\# seizures analysed for seizure duration}: number of seizures analysed for each patient based on the criteria used in this study when seizure duration was the variable of interest (response variable). \textbf{\# seizures analysed for top-delta marker of severity}: number of seizures analysed for each patient based on the criteria used in this study when top-delta marker of severity was the variable of interest (response variable).} % title of Table
\label{suppl:table_demographic} % is used to refer this table in the text
\end{sidewaystable}

% %%%%%%%%%%%%%%%%%%%%%%%%%%%%%%%%%%%%%%%%%%%%%%%%%%%%%%%%%
% %%%% Code names for subjects
% %%%%%%%%%%%%%%%%%%%%%%%%%%%%%%%%%%%%%%%%%%%%%%%%%%%%%%%%%%%%

% \begin{center}
% \begin{tabular}{||c c C||} 
%  \hline
%  & Subject & CodeName \\ [0.5ex] 
%  \hline\hline
%  1 & ID08 & ID08 \\ 
%  \hline
%  2 & ID09 & ID09 \\
%  \hline
%  3 & ID10 & ID10 \\
%  \hline
%  4 & ID14 & ID14 \\
%  \hline
%  5 & 95 & U1 \\
%  \hline
%  6 & 851 & U2 \\
%  \hline
%  7 & 934 & U3 \\
%  \hline
%  8 & 999 & U4 \\
%  \hline
%  9 & 1005 & U5 \\
%  \hline
%  10 & 1106 & U6 \\
%  \hline
%  11 & 1149 & U7 \\
%  \hline
%  12 & 1163 & U8 \\
%  \hline
%  13 & 1167 & U9 \\
%  \hline
%  14 & 1182 & U10 \\
%  \hline
%  15 & 1200 & U11 \\
%  \hline
%  16 & 1211 & U12 \\
%  \hline
%  17 & 1395 & U13 \\
%  \hline
%  18 & GLAS040 & G1 \\
%  \hline
%  19 & GLAS041 & G2 \\
%  \hline
%  20 & GLAS044 & G3 \\ [1ex] 
%  \hline
% \end{tabular}
% \end{center}

\end{document}